\documentclass[lettersize,journal]{IEEEtran}
\usepackage{amssymb}
\usepackage{amsmath}
\usepackage{graphicx}
\usepackage{caption}
\usepackage{tabularx,booktabs}
\usepackage{booktabs}
\usepackage{multirow}
\usepackage{algorithm2e}
\SetAlFnt{\small}
\SetAlCapFnt{\normalsize}
\usepackage{verbatim}
\usepackage{setspace}		
\usepackage{subfigure}
\usepackage{hyperref}
\usepackage{setspace}

\usepackage{fancyhdr}

\fancypagestyle{myheader}{
  \fancyhf{} 
  \fancyhead[C]{\vspace*{-0.5cm}\scriptsize This article has been accepted for publication in IEEE Transactions on Communications. This is the author's version which has not been fully edited\\ and
content may change prior to final publication. Citation information: DOI 10.1109/TCOMM.2024.3354204} 
}

\usepackage[style=ieee]{biblatex} 

\begin{document}

\title{RL-Based Hyperparameter Selection for Spectrum Sensing With CNNs}

\author{Amir Mehrabian,~Maryam Sabbaghian,~\IEEEmembership{Member,~IEEE}, and Halim Yanikomeroglu,~\IEEEmembership{Fellow,~IEEE}

\thanks{A. Mehrabian, and M. Sabbaghian are with the School of Electrical and Computer
Engineering, University of Tehran, Iran (e-mail: \{amir.mehrabian, msabbaghian\}@ut.ac.ir). M. Sabbaghian is the corresponding author. H. Yanikomeroglu is with the Department of Systems and Computer
Engineering, Carleton University, Ottawa, ON K1S 5B6,  Canada (e-mail: halim@sce.carleton.ca).\par
 }
}

{}

\maketitle
\thispagestyle{myheader}
\vspace{-0.7cm}

\begin{abstract}
Selection of hyperparameters in  deep neural networks is a challenging problem due to the wide search space and emergence of various layers with specific hyperparameters. There exists an absence of consideration for the neural architecture selection of convolutional neural networks (CNNs) for spectrum sensing. Here, we  develop a  method using reinforcement learning and Q-learning to systematically search and evaluate various architectures for generated datasets including different signals and channels in the spectrum sensing problem. We show by extensive simulations that CNN-based detectors proposed by our developed method  outperform  several detectors in the literature. For the most complex dataset, the proposed approach provides 9\% enhancement in accuracy at the cost of higher computational complexity. Furthermore, a novel method using multi-armed bandit model for selection of the sensing time is proposed to achieve higher throughput and accuracy while minimizing the consumed energy. The method dynamically adjusts the sensing time under the time-varying condition of the channel without prior information. We demonstrate through a simulated scenario that the proposed method improves the achieved reward by about 20\% compared to the conventional policies. Consequently, this study  effectively manages the selection of important hyperparameters for CNN-based detectors offering superior performance of cognitive radio network.

\end{abstract}


\IEEEpeerreviewmaketitle

\section{Introduction}\label{Sec1}
Having been studied for years, cognitive radios (CRs) appeared as one promising solution for underutilized spectrum. This solution gains considerable importance by bearing in mind the fast growth of spectrum usage through wireless communication \cite{Halim_new}.
 In a cognitive radio network (CRN), employing unused frequency bands of primary users (PUs) is permissible for  secondary users (SUs). A CR or an intelligent SU is primarily aimed at sensing spectrum continuously to detect idle bands for its own use and to momentarily disregard used bands. As a result, high precision spectrum sensing (SS) is  fundamental to forming a CRN.\par
Model-dependent detectors derived by likelihood ratio test (LRT) and related concepts are the basis of many earlier works in SS \cite{Axell,Taherpour}. Energy detector is the well-known and straightforward member of these detectors which is subjected to problems such as noise uncertainty, signal-to-noise ratio (SNR) wall, and limitation in handling the impulsive noise \cite{Tandra,Mehrabian, Kang2010ACO}. Consequently, various alternative methods have been proposed to overcome these issues. Recent papers on deep neural networks (DNNs) in SS showed their advantages in several areas over their model-dependent counterparts \cite{CLiu,Xie,mehrabian_new_twc}.  For instance, in SS, detectors based on convolutional neural networks (CNNs) \cite{peng,mehrabian_new} outperform the conventional model-based detectors. On the other hand,  selecting the proper architecture of network and its related hyperparameters for DNN-based detectors is a serious challenge that should be given precise consideration. In  recent works, different CNNs have been proposed for a variety of scenarios such as detection of signals with orthogonal frequency-division multiplexing  (OFDM) \cite{HLiu} and conventional signals \cite{CLiu,Xie} by use of covariance matrix (CM) of the received signal. In addition to general noise models, CNN-based detectors have shown effective performance  under the impulsive noise model as well  \cite{mehrabian_new,mehrabian_new_twc, flom-cnn}. 
 Stacked autoencoders (SAE) in \cite{QCheng} is another architecture  offered for detection of OFDM. In \cite{LSTM}, a recurrent neural network (RNN) consisted of long short-term memory (LSTM) cells has been designed for the purpose of SS. Moreover,  combinations of CNN with LSTM cells are other proposed architectures for detection in CRs \cite{mingdong,Jiandong_Xie}.\par
Although recent works in the context of CR have studied and proposed diverse networks for SS problem under distinctive signal models, this context lacks  studies on neural architecture search (NAS) method and optimization of hyperparameters in networks. Broad search space in NAS is a major challenge that requires a great amount of computational resources. In  image processing and object detection, 
search strategies  based on Bayesian optimization are a group of methods for hyperparameter optimization \cite{bergestra,domhan}. In \cite{zoph,NAS-MIT}, reinforcement learning (RL) is another approach for NAS in which an agent uses a policy \cite{NAS-MIT} or an RNN  as a controller \cite{zoph,pnas} for exploring the search space and offering a network according to the given feedback. Most of these methods are developed and tested for image classification and the related datasets. A servey of NAS methods in \cite{NasSurvey}  discusses these approaches, and their details are compactly explained. However, it is difficult to compare them  due to their completely different search spaces  and initial assumptions that each of these works considers \cite{EstebanReel}.\par
To the best of our knowledge, for the first time in this work, we  propose  a NAS method  in the context of CR and SS for datasets with samples of the received radio signals, especially for hyperparameter optimization of CNN-based detectors. This method is developed based on the Q-learning approach in \cite{NAS-MIT} to  design a suitable CNN for SS problem with different signal models.\par
In addition to network hyperparameters, sensing time is another highly influential parameter on the precision of detection, the throughput of CRN, and also the consumed energy by SU. 
In the literature, several works attempted to optimize the sensing time by assuming a time-invariant channel or by use of the channel side information (CSI) \cite{he2013,yeolLee}. Also, numerous studies tried to solve an optimization problem with energy detector (ED) using its straightforward probability statistics \cite{Haujunzhang, kong}. ED has been highly considered in SS problem due to its closed-form analytical formula for detection and false alarm probabilities, while for more complicated cases such as impulsive noise, its performance severely degrades \cite{Kang2010ACO}.
In \cite{shokri}, ED is used while a fully connected (FC) artificial neural network (ANN) controls the sensing time using  feedback of the estimated throughput.
Here, in addition to selecting proper hyperparameters of network, we also develop an RL method  for selecting the proper sensing time in CNN under non-stationary condition of channel. Although RL has been utilized for dynamic spectrum access (DSA) in CRNs \cite{Naparstek}, this work employs RL for interactive selection of sensing time. This general method is not limited to just ED and a specific wireless channel or noise model, and it requires no CSI or stationary wireless channel.\par
The key contributions of this work are as follows. We initially present a general signal model in the context of SS. Then, for the first time based on our knowledge, we develop a NAS method using Q-learning for the selection of hyperparameters of CNN-based detector for signal set of the SS problem in the presented signal model. We regard the proposed NAS method for architecture design of CNN-based detectors for Gaussian signal and OFDM signal used in IEEE 802.11 standard under different wireless channels and noise models. The proposed CNNs by the developed NAS method outperform several recent detectors in the literature. Moreover, to the best of our knowledge, a general RL-based method for adjusting  sensing time of detectors is proposed for the first time. The proposed method improves three important key parameters of throughput, detection accuracy, and  sensing computational complexity of SUs which are integrated into the designed reward function to the RL agent. Additionally, thanks to its adaptability, the method can be applied to different detectors and  action spaces, and it can be easily adapted by modifying the reward function to meet varying criteria based on specific requirements.\par
In the following, the signal model of this work is explained in Section \ref{Signalmodel}, and the proposed methods are discussed in Section \ref{Proposemethod} and Section\ref{RLforSensing}. We present simulation results of the work in Section \ref{simulation}. Finally, in Section \ref{conclusion}, the conclusion of this work is given.

\section{Signal model}\label{Signalmodel}
In an interweave cognitive network, the SU selects
a frequency channel that is presumed idle. To confirm the vacancy of the selected band, the SU senses the channel and performs a detection method to decide about the status of the sensed channel. In the case of wideband SS, multiple number of channels are sensed simultaneously with different narrowband detectors for each one \cite{Hongjian_WB_SS}. The SU with single antenna receives the following samples of baseband signal when one PU is active in the sensed band
\begin{equation}\label{eq1}
    r[n]=h[n]*s[n]+w[n].
\end{equation}
$s[n]$ is the transmitted signal by the PU which is convolved by the channel impulse response $h[n]$ and contaminated by the noise signal $w[n]$. On the other hand, when  the sensed spectrum is idle, the SU only receives noise, $r[n]=w[n]$. Now, we can model the SS problem with the following binary hypothesis testing problem \cite{kay1993fundamentals}
\begin{equation}\label{eq2}
    \left\{ \begin{array}{ll}
      H_0:r[n]=w[n]  \\
        H_1:r[n]=h[n]*s[n]+w[n]. \end{array} \right.
\end{equation}
In this hypothesis testing problem, the null hypothesis $H_0$ indicates the vacancy of the selected band and its availability for use of SU. In the  alternative hypothesis $H_1$,  PU itself is using the band, and the SU should not transmit signal at that time frame. Therefore, SS is necessary to find intervals in which the spectrum is available for the SU. To consider various possible cases, we use distinct model setups such as OFDM signal in IEEE 802.11 standard and Gaussian signal for $s[n]$, frequency selective and flat fading channels for $h[n]$, and circularly symmetric complex white Gaussian (CSCWG) and impulsive noise signals for $w[n]$. \par
In the sensing interval, the SU agent receives $N=f_sT_{sen}$ samples of the signal where $f_s$ and  $T_{sen}$ are sampling frequency and sensing time. In fact, the sensing time is a highly influential hyperparameter that can  be  interactively adjusted by the agent to improve the throughput of CR network. In the following sections, we develop  RL-based methods for selection of not only  hyperparameters of a network but also the sensing time for a more reliable and opportunistic performance of SU.

\section{Developing a Neural Architecture Search Method}\label{Proposemethod}
Although DNN-based proposed detectors have shown significant improvement in SS, the selection of architecture and hyperparameters of the network with a highly broad search space  is still a challenge, especially in the context of SS. Employing the same DNNs proposed for image classification in SS \cite{CLiu}\cite{Xie} or  using grid-search in a limited search space \cite{Oshea2018Over-the-Air} are common in the CR literature for network selection. However, with having a highly broad search space, the NAS problem is extremely complex to be solved using the grid-search method. \par
By providing datasets that include the signal activity patterns of the PUs, the proposed NAS method is able to opt effective  networks for SUs to perform SS more precisely. A computationally powerful entity within CRN can manage to select the suitable network by the NAS method, or the process can be performed in advance for finding the network architecture in that specific environment in which dataset has been acquired. Next, hyperparameters and weights of the selected and trained network are broadcasted to SUs in that specific environment.\par
Here, for NAS method, we only focus on class of CNNs and choose them for SS due to their effective performance in classification problems and downsizing the search space \cite{CLiu,mehrabian_new_twc}.
Similar to \cite{NAS-MIT}, we employ an interactive agent using RL and Q-learning algorithm \cite{watkins} to select the hyperparameters of the CNN. In this approach,  the chain-structured neural network prevents a highly complex search space for CNN, and the length of the proposed CNNs  is not fixed, and  is opted automatically by the agent, allowing agent to offer simpler networks for problems with low complexity. 

\subsection{State Parameters}
In a Markov-decision process (MDP), different state parameters encode  specific  states. The layer number ($l$) is the first state parameter which should be bounded to an upper limit to finish the generated episode. Also, the type of the layer ($t$) and parameters depending on that type are other state parameters. We selected three types of convolutional, max-pooling, and global average pooling (GAP) layers for the proposed NAS method, and these types are denoted by $t=1$, $t=2$, and $t=0$, respectively. The GAP layer computes the average output of each filter, and it was proposed in the literature to improve the network \cite{NIN-GAP}. When the agent reaches  a state with GAP layer, that state is considered as a terminal state, and episode is finished. Then, outputs of this layer, which are equal to the number of filters in the previous layer, are linked to  one FC neuron as the last layer. This final  neuron conducts the ultimate classification of the inputted signal into two classes  of pure noise ($H_0$) or PU signal with noise ($H_1$). Accordingly, number of convolutional filters (CFs), size of filters, and size of pooling window are denoted by $n_f$, $s_f$ and $s_p$, respectively, which all are  state parameters for their corresponding type of layer.\par
 For the input of the network, in-phase (I) and quadrature (Q) components of the received complex signal with length of $N$ are fed into the first layer of the network. To prevent the agent from using many pooling layers and excessive reduction of signal length, another state parameter, called pooling permission ($p_p$), is used as a flag when the length of signal is shorter than a threshold. Therefore, array  of $\textbf{s}_l=[l, t^{<l>}, n_f^{<l>}, s_f^{<l>}, p_p^{<l>}]$  with five state parameters for convolutional layers and $\textbf{s}_l=[l, t^{<l>}, s_p^{<l>}, 0, p_p^{<l>}]$ with four parameters for pooling layers encode the present state of the agent. The terminal state is reached by selection of GAP layer, and it is encoded by  $\textbf{s}_{l_t}=[l_t, 0, n_f^{<l_t-1>}, s_f^{<l_t-1>}, p_p^{<l_t-1>}]$ if the type of the previous layer is convolutional ($t^{<l_t-1>}=1$). The encoding format of the terminal state is $\textbf{s}_{l_t}=[l_t, 0, s_p^{<l_t-1>}, 0, p_p^{<l_t-1>}]$  if the previous layer is pooling ($t^{<l_t-1>}=2$).

 \subsection{Action Parameters}
The agent in each of the aforementioned states has a distinct set of options to choose its action from that set.
 Each action of the agent specifies the hyperparameters of the next layer. Also, the selected action by the agent takes the agent to a new state until the agent reaches the terminal state which finishes the episode (selection of GAP layer followed by an FC neuron at the end of the proposed network). Thus, the agent's policy and selection of actions in each state form each layer and ultimately the whole network. Options in action space of the agent $\mathcal{A}(\textbf{s}_l)$ are dependent on the current state $\textbf{s}_l$ of the agent. Different  state parameters such as type of the current layer $t$ and pooling permission $p_p$ change the possible actions in set  $\mathcal{A}(\textbf{s}_l)$ (options for selecting the hyperparameters of the next layer).\par
 A variety of different layers and activation functions can be added into the action space. Here, we regard  convolutional, max-pooling, and GAP layers for $t$.  Similarly, actions can be encoded by  array of $\textbf{a}_l=[t^{<l+1>}, n_f^{<l+1>}, s_f^{<l+1>}]$ and $\textbf{a}_l=[t^{<l+1>}, s_p^{<l+1>}]$ when type of the next selected layer is convolutional ($t^{<l+1>}=1$) and max-pooling ($t^{<l+1>}=2$), respectively. Additionally, in each state (except for the starting state with $l=0$), the agent is able to finish the episode (architecture of the network) by choosing GAP layer ($t=0$). In the initial state ($\textbf{s}_0$),  we limited the action space just to the convolutional type in order to force the agent to choose a convolutional layer for the first layer of the network and prevent the selection of max-pooling in the first layer. All parameters in the action and state space are described in the TABLE \ref{tab1n}.
 {\renewcommand{\arraystretch}{1.1}
\begin{table*}[t!]
\fontsize{9pt}{10pt}\selectfont 
  \begin{center}
    \caption{{\fontsize{9pt}{9pt}\selectfont State and action parameters.}  \label{tab1n}}
    \begin{tabular}{>{\centering}m{.12\textwidth}|>{\centering}m{.5\textwidth}|>{\centering}m{.12\textwidth}
    |>{\centering\arraybackslash}m{.12\textwidth}} 
  \toprule[1.5pt]
  \textbf{Parameters}&\textbf{Discriptions}&\textbf{ State  Parameters} &   \textbf{Action Parameters}\\
\toprule[1.5pt]

   $ l$& \raggedright Layer index &$l$& $t^{<l>}$\\
     $t^{<l>}$& \raggedright Type of the $l$-th layer&$t^{<l>}$&$n_f^{<l>}$\\
     $n_f^{<l>}$&\raggedright  Number of CFs in the $l$-th layer (if $t^{<l>}=1$) &$n_f^{<l>}$&$s_f^{<l>}$\\
     $s_f^{<l>}$& \raggedright  Size of CFs in the $l$-th layer (if $t^{<l>}=1$)& $s_f^{<l>}$& $s_p^{<l>}$\\
    $s_p^{<l>}$& \raggedright Size of pooling in the $l$-th layer (if $t^{<l>}=2$)& $s_p^{<l>}$&\\
     $p_p^{<l>}$&\raggedright  Pooling permission& $p_p^{<l>}$&\\
      \toprule[1.5pt]
      
    \end{tabular}
  \end{center}
  \vspace{-0.7cm}
\end{table*}
}
 \subsection{Episodes and Rewards}
 After action $\textbf{a}_l$ was adopted by the agent in state $\textbf{s}_l$, that agent moves to the next state with an array of $\textbf{s}_{l+1}$. In the next state array, $l$ is incremented, and other parameters are updated based on the selected action
 $\textbf{a}_l$. If $t=2$ (the $l$-th layer is max-pooling with pooling size of $s_p$), and  the length of signal in the output of the layer is shorter than $N_{trsh}$, the pooling permission flag $p_p$ should be toggled to zero (no more pooling layer is permitted in the network). If this layer is not max-pooling or the length of the output signal is not excessively short, this flag remains one. Therefore, the new state can be determined in parameters of $\textbf{s}_{l+1}$. The agent chooses actions and moves to new states. The episode continues until the agent selects GAP layer ($t=0$), or the number of layers in the network reaches  its limit ($L$) where the  agent has  to choose GAP layer and finish the episode. \par
 By the end of an episode, a string of states, actions, and rewards ($\textbf{s}_0$, $\textbf{a}_0$, $r_0$, ... ) is provided. Based on these episodes, we manage to estimate the action-value functions $Q([\textbf{s}_l,\textbf{a}_l])$. If an accurate estimation of $Q([\textbf{s}_l,\textbf{a}_l])$ can be computed, accordingly, the agent selects the optimal actions (layers) in each state to form a suitable network \cite{sutton2018reinforcement}. 

 For SS, accuracy of classification of spectrum into $H_0$  (the idle frequency band) and $H_1$ (the frequency band is used by PU) is the key feature that specifies the quality of the CNN-based detector. We aim to select hyperparameters of CNN in order to guarantee that the  selected network offers the highest classification accuracy. Thus, we associate the achieved reward by the agent with   the probability of correct classification  ($P_c$) for the generated network on  a specific dataset. Since the reward is based on the $P_c$ for the generated episode, we should wait until the end of the episode. When a network is generated, and the agent reaches  the terminal state, it achieves its terminal reward, $r_{l_t}=P_c$. In other words, we can evaluate the taken actions by the agent at the end of the episode based on the entire performance of the generated network. Accordingly, no reward is given to the agent after each action (selection of one layer). For this reason, we award the agent by this reward function $r_l = P_c\delta(l-l_t)$ where $\delta(.)$ is the delta function.
Therefore, the proposed  NAS method  attempts to offer architectures with the highest classification accuracy for the specified dataset. 

 \subsection{Updating Action-Value Functions}
 The agent uses each generated episode to estimate action-value functions $Q([\textbf{s}_l,\textbf{a}_l])$ for various $\textbf{s}_l$ and $\textbf{a}_l$ in different layers ($l$) of the network. It is important for the agent to achieve an accurate estimation of $Q([\textbf{s}_l,\textbf{a}_l])$ with the least possible number of generated episodes. The proposed NAS method should be conducted by an administrative entity with high computational resources in CRN for a specific dataset  with  received signals in a special environment. Then, the information of the best network with its proper amount of weights for that special signal set in that specific environment can be transferred to SUs for SS. Finding an accurate estimation of $Q([\textbf{s}_l,\textbf{a}_l])$ with the least possible number of  generated episodes is required  to expedite the NAS method and reduce the computational complexity. The agent needs to select actions based on an adequate probability distribution ($\mathrm{Pr}[\textbf{a}_l|\textbf{s}_l]$), called policy and usually denoted by $\pi([\textbf{s}_l,\textbf{a}_l])$. For taking actions in each state ($\textbf{s}_l$), this policy should   maintain a proper balance between exploration and exploitation to avoid generation of excessive episodes. Generally, after the initial exploration phase with sufficient number of episodes, exploitative policies assist the agent to concentrate on taking actions with higher ranks based on their $Q([\textbf{s}_l,\textbf{a}_l])$. This also increases the precision of action-value function of possible optimal actions, $\textbf{a}_l^*$, by repetition of visiting states with higher values and taking  actions whose ranks are higher in accordance to their  $Q([\textbf{s}_l,\textbf{a}_l])$.\par
 Similar to \cite{NAS-MIT}, we use Q-learning  with $\epsilon$-greedy policy to adjust the exploration and exploitation phases of the agent's policy. Q-learning is a technique employed to systematically explore the problem's search space, aiming to identify the optimal sequence of actions. The key advantage of utilizing Q-learning lies in its ability to perform a systematic search and learn from the experience by updating $Q([\textbf{s}_l,\textbf{a}_l])$ based on the outcomes of actions taken in the environment. This results in a significant reduction in the number of required samples and accelerates the search process. Additionally, Q-learning is an off-policy method where the agent updates the estimation of action-value functions based on the greedy policy, called target policy, while it takes actions by use of another policy, called behavior policy, which is commonly an exploratory policy (soft policy) similar to $\epsilon$-greedy with a  large $\epsilon$ \cite{sutton2018reinforcement}. For Q-learning, the updating rule for action-value functions is \cite{watkins}
 \begin{equation}\label{eq4}
  \begin{split}
    Q([\textbf{s}_l,\textbf{a}_l])& \leftarrow   Q([\textbf{s}_l,\textbf{a}_l]) + \\
     &\alpha\left[r_l +\gamma_d \max_{\textbf{a}_{l+1}}Q([\textbf{s}_{l+1},\textbf{a}_{l+1}])-Q([\textbf{s}_l,\textbf{a}_l])\right],
\end{split}
\end{equation}
where $\gamma_d$ is the discount factor, and $\alpha$ is the step size. The maximum operation  in the updating rule indicates the greedy target policy. Thus,  each action-value function is updated by assumption of the greedy selection of action in the next state. The agent receives a reward and uses (\ref{eq4}) for updating all action-value functions  from the terminal layer, $l=l_t$, to the first layer in the episode. \par
For updating $Q([\textbf{s}_{l_t-1},\textbf{a}_{l_t-1}])$ (state-actions before the terminal states, $\textbf{s}_{l_t}$) only reward is effective since action-value function of its next state, which is the terminal state, is set to zero, $Q([\textbf{s}_{l_t},-])=0$.
Additionally, we need a far-sighted agent since only accuracy of networks $P_c$, at the end of episodes is important for SS. Thus, discount factor is set to zero ($\gamma_d=0$). Moreover, we assume the problem is stationary since the dataset used for finding $P_c$ is fixed during the generation of episodes and the whole process.  Consequently, the step size in the updating rule of (\ref{eq4}) is not constant. It is chosen for each action-value function with $\alpha = 1/N([\textbf{s}_l,\textbf{a}_l])$ formula where
$N([\textbf{s}_l,\textbf{a}_l])$ is equal to  the number of selection of that action-state. In contrast to \cite{NAS-MIT} which employed a constant step size, this variable decreasing step size for estimation of $Q([\textbf{s}_l,\textbf{a}_l])$ in (\ref{eq4}) is advantageous in a stationary problem. In fact, a variable decreasing step size satisfies the required conditions of Robbins-Monro for the convergence of the estimated action-value functions to their true values in the updating rule \cite{sutton2018reinforcement}.

\subsection{The Developed Algorithm for NAS}\label{subsect3e}
{\fontsize{9pt}{8pt}\selectfont
\RestyleAlgo{ruled}
\begin{algorithm}[t!]

\caption{Algorithm of the proposed NAS method \label{algor1}
  \vspace{-0.35cm}}
 
\DontPrintSemicolon

\KwIn{
States and actions in MDP, $L$, $N_{trsh}$, $N_{epi}$, dataset}

        \Begin(Initialization of parameters $\forall\  [\textbf{s}_l,\textbf{a}_l]$ in MDP:)
            {$Q([\textbf{s}_l,\textbf{a}_l])\leftarrow$ arbitrary\;
             $Q([\textbf{s}_{l_t},-])\leftarrow$ 0\;
             $N([\textbf{s}_l,\textbf{a}_l])\leftarrow 0 $\;
        }
        \For {$n_{epi}=1$ to $N_{epi}$}
        {
             Choose a proper $\epsilon$ \;
            \Begin( Generation of an episode with a string of $\textbf{s}_0$, $\textbf{a}_0$, $r_0$, ..., $\textbf{s}_{l_t}$, ${r}_{l_t}$:)
                {
                 Start from state $\textbf{s}_0$\;
                 \For {$l=0$ to $l_t-1$}
                     {
                    \Begin(Taking action $\textbf{a}_l$ based on $\epsilon$-greedy:)
                    {
                       $\textbf{a}_l^{*} \leftarrow  \max_{\textbf{a}_{l}}Q([\textbf{s}_{l},\textbf{a}_{l}])$\;
                        $\pi([\textbf{s}_{l},\textbf{a}_{l}]) =\left\{ \begin{array}{ll}
                         1-\epsilon +\epsilon/|\mathcal{A}(\textbf{s}_l)|, &\textbf{a}_{l}= \textbf{a}^{*}_{l} \\
                        \ \epsilon/|\mathcal{A}(\textbf{s}_l)|, &\textbf{a}_{l}\neq \textbf{a}^{*}_{l} \end{array} \right.$\;}
                        Receive reward $r_l$ \;
                       Move to $\textbf{s}_{l+1}$\;

                       }
                 Generate network and use the dataset and K-fold validation for finding $P_c$ \;
                 Receive reward $r_{l_t}=P_c$\;}
          \Begin(Update action-value functions in the generated string from the terminal state to the initial state:)
          {
                \For {$l=l_{t}-1$ to $0$}
                {
                      $Q([\textbf{s}_l,\textbf{a}_l])\leftarrow Q([\textbf{s}_l,\textbf{a}_l]) + \frac{1}{N([\textbf{s}_l,\textbf{a}_l])}\left[r_l + \max_{\textbf{a}_{l+1}}Q([\textbf{s}_{l+1},\textbf{a}_{l+1}])-Q([\textbf{s}_l,\textbf{a}_l])\right]$\;
                     $N([\textbf{s}_l,\textbf{a}_l])\leftarrow N([\textbf{s}_l,\textbf{a}_l])+1$\;
                }
           }
         }
 \KwRet{\  } { $Q([\textbf{s}_l,\textbf{a}_l])$ and $N([\textbf{s}_l,\textbf{a}_l])$ $\ \forall \ [\textbf{s}_l,\textbf{a}_l]$ in MDP}
 
 \end{algorithm}

 }

The proposed NAS method to select hyperparameters of CNN-based detectors for SS datasets is explained in Algorithm \ref{algor1}. Here, we limit the maximum number of layers in the generated networks to $L$, and the minimum length of the samples of signals after pooling layers to $N_{trsh}$. Then, we generate proper number of episodes equal to $N_{epi}$ in order to find an accurate estimation of $Q([\textbf{s}_l,\textbf{a}_l])$ for all state-actions. The  algorithm aims to find the best CNN architectures for SS in the specified CR network in which the dataset is acquired. Clearly, different signal datasets may result in  different offered network architectures by the algorithm.\par
 One can arbitrarily pass the initial values for the action-value functions at the beginning of the algorithm. For the long-term required update of the offered network's architecture, the algorithm can be rerun with the final estimation of $Q([\textbf{s}_l,\textbf{a}_l])$ in the previous run  as initial values for the new run to prevent the repetition of the process.
 For each generated episode, a simple K-fold validation is conducted on the trained network, and the mean accuracy of the network for all folds, $P_c$, is returned as $r_{l_t}$. The final output of the algorithm is a bank of all action-value functions where the best suggested network can be found by choosing actions with the highest $Q([\textbf{s}_l,\textbf{a}_l])$  for each state in this bank. 
\section{RL for Selecting Sensing Time}\label{RLforSensing}

\begin{figure}[t!]
\centering
		\includegraphics[width=3.3in]{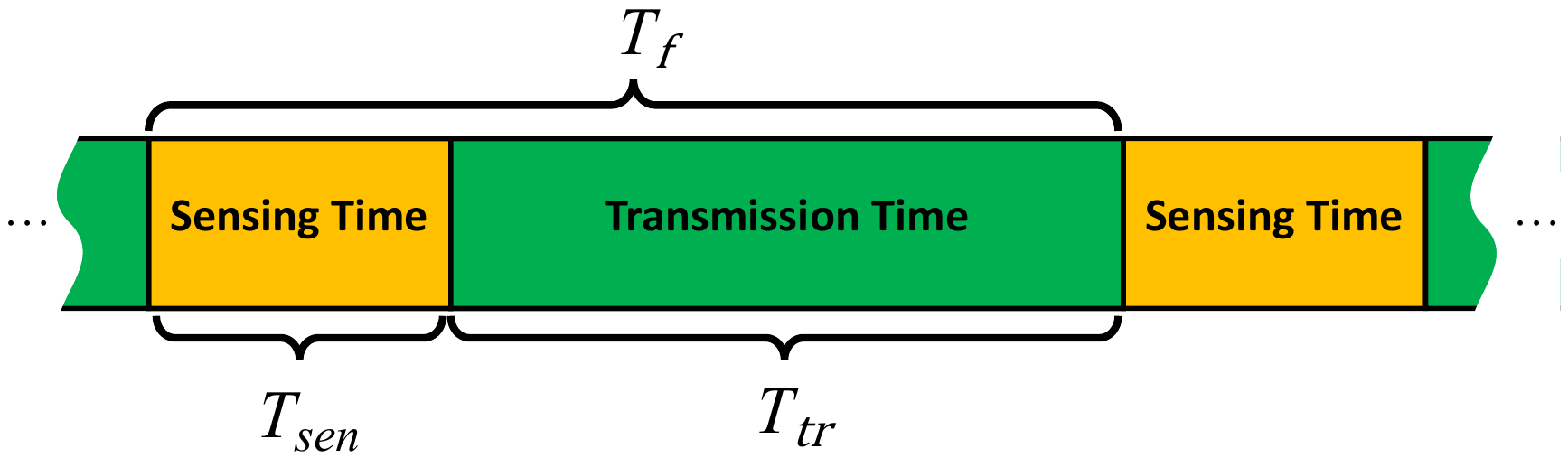}
\centering	
	\caption{{\fontsize{9pt}{10pt}\selectfont Time frames with sensing and transmission times for SU.}\label{fig3}} 
	\vspace{-0.7cm}
\end{figure}

Another important hyperparameter of DNN-based detectors is the number of samples in the  signal inputted to the network. In most cases, a time frame is regarded with the duration of $T_{f}$ divided to a sensing interval $T_{sen}$ and a transmission interval $T_{tr}=T_f-T_{sen}$. The SUs only sense the channel frequency and avoid transmitting during the sensing interval. If a sensed channel is regarded idle, the rest of the time frame is employed by the SU for the transmission of its own message. For multiple SUs, utilizing either centralized coordination and guaranteed access or random access can be effective in minimizing collisions during transmission \cite{Qian_MAC}. On the other hand, the PU is active, and the SU is not permitted to transmit during $T_{tr}$. Fig. \ref{fig3} illustrates the related time frames and their division into different sections.\par
 The allocated detector for that frequency band senses the band all the time to transmit its data when that specific band is idle. This is similar to the case where the wideband spectrum is divided into several narrowband channels, and a separate detector is utilized for each of these narrowband channels which is similar to the case of filter-bank detection \cite{Hongjian_WB_SS,Bashar-wideband}.\par
Sensing time is an influential parameter on the accuracy of detection, the throughput of the CR network, and the consumed energy by SU for sensing. Adjusting sensing time dynamically under various conditions can improve the entire performance of the CR network. Indeed, in a low-SNR regime, accurate detection of PU's activity is difficult, and a longer sensing duration results in a lower probability of misdetection. This implies that SUs misinterpret an occupied channel as an available channel for their transmission with a lower likelihood. Thus, SUs create less interference for PU's communication. However, longer sensing time shortens the transmission time for SU to transmit its data. Alternatively, in case of a PU transmitting signals with a large SNR, a small duration of sensing suffices for the detection of signal by SU due to the large SNR of signal. By this means, SU can stop wasting its computational and time resources for a long sensing time or sensing that specific active  channel. Therefore, it is desired that a smart SU can adjust its sensing time based on the condition of wireless channel  and the received power of PU's signal to the SU. \par
To achieve this, here,  SU acts as an agent and uses an algorithm based on RL to dynamically adjust the sensing time. In contrast to the previous section, we consider this problem with a single state as a multi-armed bandit (MAB) where this simpler single-state modeling requires fewer number of average reward estimation and sampling. SU (agent) should take action $a$ from the action space $\mathcal{A}$. Actions are related to the length of the sensing time, and consequently, the number of samples of the inputted signal to the network. Accordingly, SU is equipped with $|\mathcal{A}|$ number of networks or detectors' setup. If SU  selects action $a$, which offers a long sensing time and large $N$, a detector with proportionate input size of $N$ is used for spectrum sensing. This detector obtains a higher probability of detection in comparison to a detector with a shorter signal length. However, in this case, SU requires more computational complexity, and it has a shorter transmission time. Based on physical environment, wireless channel, location of SU and PU, and power of the received signal, SU  is able to adjust its sensing time interactively by implementing this proposed method. In spite of several works in the literature,  our method enables the agent to adjust the sensing time even without CSI \cite{he2013}.
\begin{figure}[t!]
\centering
		\includegraphics[width=3.3in]{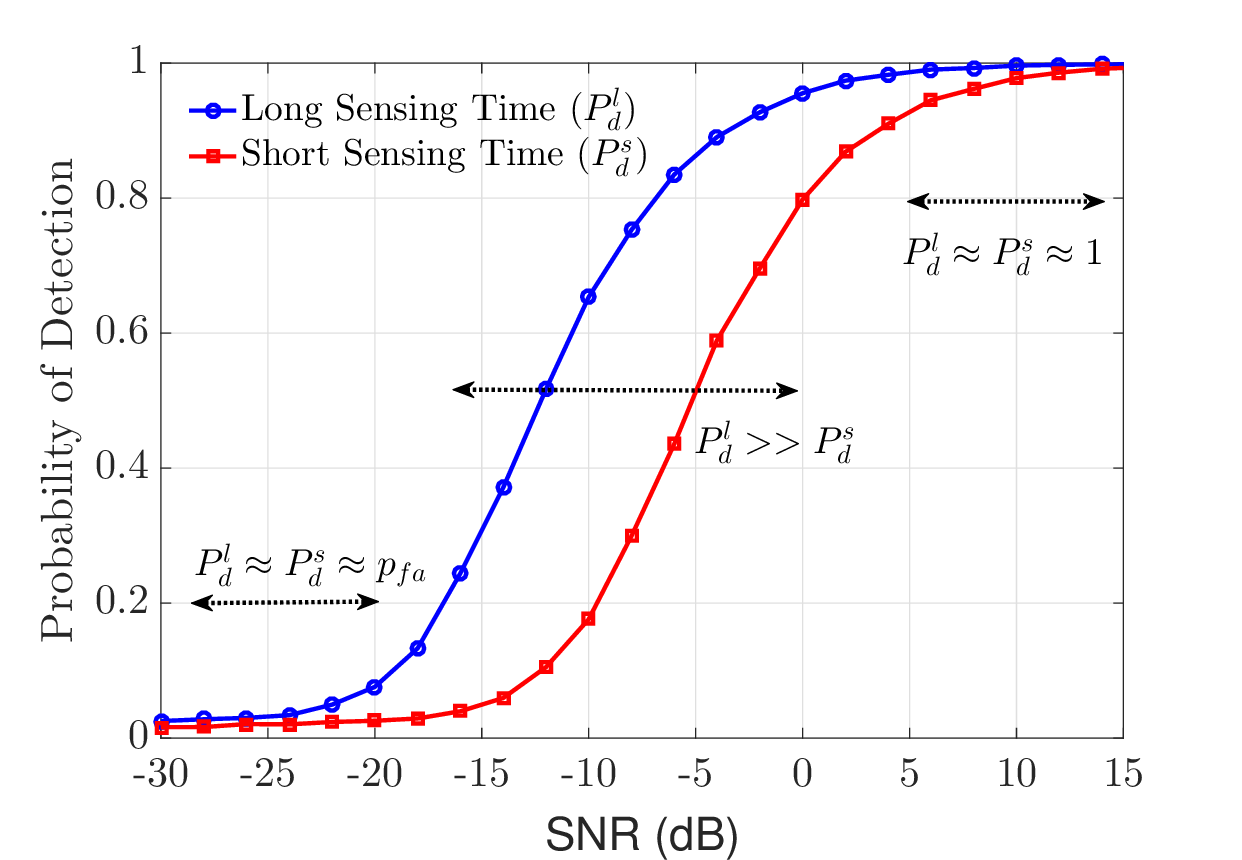}
\centering	
	\caption{{\fontsize{9pt}{10pt}\selectfont Probability of detection of  detectors with long and short sensing times for three intervals of SNR.}\label{fig4}} 
	\vspace{-0.5cm}
\end{figure}
For rewarding the agent, various scenarios should be regarded based on the decision of detector $D_i$  about the condition of the sensed channel ($D_0$ and $D_1$ mean that SU decided that the channel is vacant or full, respectively), and its true condition ($H_0$ or $H_1$). Consequently, probability of two important events of correct detection and false alarm can be referred to as probability of  detection $P_d=\mathrm{Pr}[D_1|H_1]$ and probability of false alarm  $P_{fa}=\mathrm{Pr}[D_1|H_0]$, respectively.\par
Here, without loss of generality,  we only consider two actions of long and short sensing times ($a=l$ or $a=s$) for the sake of simplicity in the mathematical analysis of different conditions.
Fig. \ref{fig4} provides interesting intuition about probability of detection for long and short sensing time actions versus SNR. As seen, for a wide range of SNRs, probability of detection for the long sensing time, $P_d^l(\mathsf{SNR})=\mathrm{Pr}[D_1^l|H_1]$ (the blue curve), is greater than that of the short sensing time, $P_d^s(\mathsf{SNR})=\mathrm{Pr}[D_1^s|H_1]$ (the red curve). In a low-SNR regime, both the short and long sensing time actions (we refer to them as long and short actions in the following) result in $P_d^l(\mathsf{SNR})\approx P_d^s(\mathsf{SNR})\approx p_{fa}$ where $p_{fa}$ is the desired probability of false alarm (the left interval). Generally, it is desired for a detector to have a false alarm probability less than the desired false alarm upper bound $P_{fa} \le p_{fa}$, which can be possible by choosing a proper detection threshold. For medium SNRs, long action offers a higher detection probability $P_d^l(\mathsf{SNR})>> P_d^s(\mathsf{SNR})$ (the middle interval), and when SNR is quite large, we expect $P_d^l(\mathsf{SNR})\approx P_d^s(\mathsf{SNR})\approx 1$ (the right interval). \par
By considering $H_i$ and $D_j$ for both long and short actions, eight scenarios are possible shown in  Table \ref{tab1} with the related reward ($r^a_{i,j}$), true state ($H_i$), the decided state for the sensed channel by detector ($D^a_j$), the related probability, throughput (T), possible interference for PU (I), and complexity (C). Here, probability of state of channel is denoted by $P_i=\mathrm{Pr}[H_i]$.\par 
For the first and second scenarios, the sensed spectrum is vacant, and both long and short actions correctly recognize that with the same probability. This happens since we assumed that in $H_0$, detectors are adjusted to offer a  probability of false alarm  approximately equal to the desired probability of false alarm $P_{fa}\approx p_{fa}$. An agent (SU) with long and short actions has the opportunity of transmitting with rate of $R_{SU}$ in  the transmission times of $T_{tr}^s$ and $T_{tr}^l$, respectively. In all scenarios, the complexity of SS is considered proportionate to the sensing time of each action $T^{a}_{sen}$. In the cases described in rows 3 and 4 of Table \ref{tab1}, both actions missed the opportunity of the vacant spectrum.\par When the true state of the sensed channel is $H_1$, and misdetection occurs, SU transmits its signal which creates interference for PU's communication (rows 5 and 6 in Table \ref{tab1}). Then, due to the interference, we assume that SU receives no acknowledgment from the receiving node in CR network \cite{he2013,Naparstek}. Thus, SU recognizes its misdetection and is able to punish the agent. \par 
{\renewcommand{\arraystretch}{1.2}
\begin{table*}[tb!]
\fontsize{9pt}{9pt}\selectfont 
  \begin{center}
    \caption{{\fontsize{9pt}{10pt}\selectfont Various scenarios of reward for actions of the agent.}  \label{tab1}}

    \begin{tabular}{>{\centering}m{.03\textwidth}
    >{\centering}m{.09\textwidth}
>{\centering}m{0.09\textwidth}
 >{\centering}m{0.09\textwidth}
 >{\centering}m{0.2\textwidth}
 >{\centering}m{0.06\textwidth}
  >{\centering}m{0.06\textwidth}
    >{\centering\arraybackslash}m{0.06\textwidth}} 
  \toprule[1.5pt]

&\textbf{Reward} &  \textbf{True State} & \textbf{Decided State} & \textbf{Probability of Event} &\textbf{ T}& \textbf{I}& \textbf{C}\\
\toprule[1.5pt]
      \textbf{}&$r_{i,j}^a$&$H_i$& $D^a_j$&$P_{i,j}^a$&$T_{i,j}^a$&$I_{i,j}^a$&$C_{i,j}^a$\\
      \toprule[1.5pt]
(1)    &  $r_{0,0}^s$ &$H_0$ & $D_0^s$& $P_0(1-p_{fa})$   &$T^s_{tr}R_{SU}$ & -& $T^s_{sen}$\\
      \hline
     (2)        & $r_{0,0}^l$ &$H_0$ & $D_0^l$& $P_0(1-p_{fa})$   &$T^l_{tr}R_{SU}$  & -& $T^l_{sen}$\\
      \hline
     (3)    & $r_{0,1}^s$ &$H_0$ & $D_1^s$& $P_0p_{fa}$   &- & -& $T^s_{sen}$\\
      \hline
    (4)    &  $r_{0,1}^l$ &$H_0$ & $D_1^l$& $P_0p_{fa}$ &- & -& $T^l_{sen}$\\
      \hline
    (5)    &  $r_{1,0}^s$ &$H_1$ & $D_0^s$& $P_1[1-P_{d}^s(\mathsf{SNR})]$   &- & $\xi$ &$T^s_{sen}$\\
      \hline
    (6)    &  $r_{1,0}^l$ &$H_1$ & $D_0^l$& $P_1[1-P_{d}^l(\mathsf{SNR})] $  &- & $\xi$&$T^l_{sen}$\\
      \hline
    (7)    &  $r_{1,1}^s$ &$H_1$ & $D_1^s$& $P_1P_{d}^s(\mathsf{SNR})$   &- & -& $T^s_{sen}$\\
      \hline
     (8)    & $r_{1,1}^l$ &$H_1$ & $D_1^l$&  $P_1P_{d}^l(\mathsf{SNR})$    &- & -& $T^l_{sen}$\\
      \toprule[1.5pt]
    \end{tabular}
  \end{center}
  \vspace{-.5cm}
\end{table*}
}
For these  scenarios, a corresponding reward is defined, $r_{i,j}^a$. We assume that the top  priority of a SU is to avoid creating interference for PUs which is directly related to $P_d$ and the sensing time. In a lower rank of priority, SU needs to find and use the vacant channels with the least possible complexity. Therefore, in order to force the agent to follow these priorities, we define the rewards as a weighted sum of throughput, interference, and complexity. This reward is formulated as $r_{i,j}^a=\lambda_{1}T^{a}_{i,j}-\lambda_{2}I^{a}_{i,j}-\lambda_{3}C^{a}_{i,j}$ where $\lambda_{2}I^{a}_{i,j}>>\lambda_{1}T^{a}_{i,j}>>\lambda_{3}C^{a}_{i,j}> 0$, for all $i$, $j$, and $a$ except for cases in which $I^{a}_{i,j}$ and $T^{a}_{i,j}$  are equal to zero. However, to force the agent to act differently, one can set different weights to change the priority list of the agent.\par  It is interesting to follow the expected reward $Q(a)=\mathrm{E}\{r_{i,j}^a\}$ attained by the agent for both short and long actions in the following cases to understand its behavior for gaining the maximum average reward (the least interference and sensing complexity together with  the highest possible throughput).

\subsubsection*{Case1} Let us assume the true state as $H_0$ and $P_0=P_1$. Using rows 1 and 3 in Table \ref{tab1}, the expected reward for short action is 
\begin{equation}\label{eq5}
\begin{split}
   \mathrm{E}&\{r_{0,j}^s\}=Q(s|H_0) =(1-p_{fa})r_{0,0}^s+p_{fa}r_{0,1}^s=\\
   &(1-p_{fa})(\lambda_{1}T_{tr}^sR_{SU}-\lambda_{3}T_{sen}^s)-p_{fa}\lambda_{3}T_{sen}^s,
   \end{split}
\end{equation}
and for long action by using rows 2 and 4 in table \ref{tab1}, we have
\begin{equation}\label{eq6}
\begin{split}
   \mathrm{E}&\{r_{0,j}^l\}=Q(l|H_0) =(1-p_{fa})r_{0,0}^l+p_{fa}r_{0,1}^l=\\
   &(1-p_{fa})(\lambda_{1}T_{tr}^lR_{SU}-\lambda_{3}T_{sen}^l)-p_{fa}\lambda_{3}T_{sen}^l.
   \end{split}
\end{equation}
For each action, a different detector with longer or shorter length of the input signal is used ($T_{sen}^l>T_{sen}^s$, and consequently, $T_{tr}^l<T_{tr}^s$). However, in equations (\ref{eq5}) and (\ref{eq6}), we use the fact that detection threshold of both networks are adjusted in the way that at the worst case under $H_0$, probability of false alarm for both networks become $P_{fa}\approx p_{fa}$. By comparing the expected reward of both actions in (\ref{eq5}) and (\ref{eq6}), we conclude that $Q(s|H_0)>Q(l|H_0)$. It means  that under $H_0$, short sensing time is the optimum action for the agent.

\subsubsection*{Case2} For the following cases, we consider $H_1$ with different SNRs for the received signal in the SU. In this case, we assume SNR is considerably low, similar to the left interval of Fig. \ref{fig4} where $\mathsf{SNR}\in \{\mathsf{SNR}|P_d^s(\mathsf{SNR})\approx P_d^l(\mathsf{SNR})\approx p_{fa}\}$. Thus, the expected reward for short action  in rows 5 and 7 of Table \ref{tab1} is
\begin{equation}\label{eq7}
\begin{split}
   \mathrm{E}&\{r_{1,j}^s\}=Q(s|H_1) =[1-P_d^s(\mathsf{SNR})]r_{1,0}^s+P_d^s(\mathsf{SNR})r_{1,1}^s\\
   &\approx(1-p_{fa})(-\lambda_{2}\xi-\lambda_{3}T_{sen}^s)-p_{fa}\lambda_{3}T_{sen}^s,
   \end{split}
\end{equation}
and for long action in rows 6 and 8 of Table \ref{tab1}, it is equal to 
\begin{equation}\label{eq8}
\begin{split}
   \mathrm{E}&\{r_{1,j}^l\}=Q(l|H_1) =[1-P_d^l(\mathsf{SNR})]r_{1,0}^l+P_d^l(\mathsf{SNR})r_{1,1}^l\\
   &\approx(1-p_{fa})(-\lambda_{2}\xi-\lambda_{3}T_{sen}^l)-p_{fa}\lambda_{3}T_{sen}^l.
   \end{split}
\end{equation}
These expected rewards in (\ref{eq7}) and (\ref{eq8}) show that the punishment for action of short sensing time is slightly smaller than long sensing time.
\subsubsection*{Case3}
In this case, again, we assume $H_1$ and the medium SNRs (the middle interval of Fig. \ref{fig4}) where $\mathsf{SNR}\in \{\mathsf{SNR}|P_d^s(\mathsf{SNR})<< P_d^l(\mathsf{SNR})\}$. The formulation of the expected reward for both actions is similar to (\ref{eq7}) and (\ref{eq8}) with different probabilities of detection. Thus,  for short and long actions, we conclude that
\begin{equation}\label{eq9}
   Q(s|H_1) =[1-P_d^s(\mathsf{SNR})](-\lambda_{2}\xi-\lambda_{3}T_{sen}^s)-P_d^s(\mathsf{SNR})\lambda_{3}T_{sen}^s,
\end{equation}
\begin{equation}\label{eq10}
   Q(l|H_1) =[1-P_d^l(\mathsf{SNR})](-\lambda_{2}\xi-\lambda_{3}T_{sen}^l)-P_d^l(\mathsf{SNR})\lambda_{3}T_{sen}^l.
\end{equation}
Since $\xi>>T_{sen}^l>T_{sen}^s$, the amount of the first term on the right side of equations (\ref{eq9}) and (\ref{eq10}) dominates the whole term and the received punishment by the agent. Consequently, because of a smaller chance of misdetection for long action ($1-P_d^s(\mathsf{SNR})>>1-P_d^l(\mathsf{SNR})$), this action causes less interference  for PU in various trials, and the agent receives a smaller punishment. Therefore, in this case, long action is the optimum one.
\subsubsection*{Case4} Finally, we regard large SNRs where $\mathsf{SNR}\in \{\mathsf{SNR}|P_d^s(\mathsf{SNR})\approx P_d^l(\mathsf{SNR})\approx1\}$  presents the right interval in Fig. \ref{fig4}. Therefore, we can rewrite the  equations of (\ref{eq7}) and (\ref{eq8}) as  $Q(s|H_1) \approx -\lambda_{3}T_{sen}^s$, and  $Q(l|H_1) \approx -\lambda_{3}T_{sen}^l$. It is clear  when the amount of SNR is very large, both detectors decide correctly, while short action is the best action since it conserves more energy for SU due to smaller required computation. All these results are briefly reported in Table \ref{tab2} with the best action for each case.\par
In the following section, $\epsilon$-greedy and gradient bandit (GB) policies \cite{sutton2018reinforcement} are employed for the agent to choose the best sensing time under each case. Similar to Subsection \ref{subsect3e}, we utilize
\begin{equation}\label{eq11}
   \pi(a) =\left\{ \begin{array}{ll}
                      1-\epsilon +\epsilon/|\mathcal{A}|, &a
                     = a^{*} \\
                    \ \epsilon/|\mathcal{A}|, &a\neq a^{*},
                    \end{array} \right.
\end{equation}
where $a\in\mathcal{A}$ and  $a^{*} =\max_{a}\hat{Q}(a)$, and for estimating $Q(a)$, this updating rule is applied
\begin{equation}\label{eq12}
\hat{Q}(a)\leftarrow \hat{Q}(a)+ \alpha_{lr}\left[r_{i,j}^{a}-\hat{Q}(a)\right].
\end{equation}
$\hat{Q}(a)$ is the estimation of the expected reward $Q(a)$, and $\alpha_{lr}$ is the step size.\par

For GB, the policy  follows a soft-max distribution \cite{sutton2018reinforcement}
\begin{equation}\label{eq13}
  \mathrm{Pr}[a]= \pi(a) =\frac{e^{H(a)}}{
   \sum_{b\in\mathcal{A}}e^{H(b)} }\ ,
\end{equation}
while it alters the selection probability  of each action based on the preference of action $H(a)$. The policy changes during trials in favor of actions with higher preferences. In fact, if $a$ is the selected action, the following equation shows that actions have higher preferences when they receive rewards  greater than a benchmark
\begin{equation}\label{eq14}
   \left\{ \begin{array}{ll}
  H(a) \leftarrow H(a)+ \alpha_{pr}(r_{i,j}^a-\hat{Q}(a))[1-\pi(a)],  \\
 H(b) \leftarrow H(b)- \alpha_{pr}(r_{i,j}^a-\hat{Q}(b))\pi(b),\forall b \in \mathcal{A}-\{a\}.
                      \\ \end{array} \right.
\end{equation}
This benchmark is $\hat{Q}(a)$ in (\ref{eq12}), and $\alpha_{pr}$ is the step size for the preference of actions. In contrast to the NAS method, we will set $\alpha_{lr}$ and $\alpha_{pr}$ with small constant amounts since this problem is non-stationary \cite{sutton2018reinforcement} due to the change of active and idle frequency channels. This helps the agent quickly modify its policy in favor of action with the highest reward when the state of the sensed channel ($H_i$) or the received SNR changes.
{\renewcommand{\arraystretch}{1.1}
\begin{table}[tb!]
\fontsize{9pt}{10pt}\selectfont 
  \begin{center}
    \caption{{\fontsize{9pt}{9pt}\selectfont The best sensing time for the agent under various conditions.} \label{tab2}}

    \begin{tabular}{
 >{\centering}m{0.05\textwidth}
 >{\centering}m{0.12\textwidth}
  >{\centering}m{0.1\textwidth}
    >{\centering\arraybackslash}m{0.13\textwidth}} 
  \toprule[1.5pt]

\textbf{Case} &  \textbf{True State} & \textbf{SNR} & \textbf{Best Action}\\
\toprule[1.5pt]
(1)    &  $H_0$ & - & Short ($s$)\\
      \hline
     (2)        & $H_1$ & Low & Short ($s$) \\
      \hline
     (3)    & $H_1$ & Medium & Long ($l$)\\
      \hline
    (4)    & $H_1$ & High & Short ($s$)\\
      \toprule[1.5pt]
    \end{tabular}
  \end{center}
  \vspace{-0.8cm}
\end{table}
}%

{\renewcommand{\arraystretch}{1.1}
\begin{table*}[tb!]
\fontsize{9pt}{10pt}\selectfont 
  \begin{center}
    \caption{{\fontsize{9pt}{9pt}\selectfont The best offered CNN architecture by the proposed NAS method for datasets.} \label{tab3}}

    \begin{tabular}{
 >{\centering}m{0.07\textwidth}
 >{\centering}m{0.09\textwidth}
  >{\centering}m{0.11\textwidth}
  >{\centering}m{0.08\textwidth}
  >{\centering}m{0.26\textwidth}
   >{\centering}m{0.12\textwidth}
    >{\centering\arraybackslash}m{0.08\textwidth}} 
  \toprule[1.5pt]

\textbf{Dataset} &  \textbf{Signal} & \textbf{Noise/ Channel} &\textbf{Sensing Time}&     \textbf{NAS-CNN}& \textbf{Run Time (GPU days)} & \textbf{Mean Training Time} \\
\toprule[1.5pt]
(1)    &  Gaussian & CSCWG\\ Flat Fading &100 samples &Conv(64,3), GAP & 4.46 & 0.48 s\\
      \hline
     (2)    &  OFDM & S$\alpha$S \\ EPA & 8 \textmu s  &Conv(64,3), Conv(64,3), Conv(32,5), Conv(32,5), Conv(16,3), Conv(16,3), GAP & 7.238 & 0.74 s\\
      \hline
     (3)   &  OFDM & S$\alpha$S\\ EPA & 32 \textmu s  &Conv(32,3), Conv(32,3), Conv(64,5), Conv(64,5), Conv(16,5), Conv(16,5), Conv(8,3), Conv(64,3), GAP & 10.947& 2.42 s\\

      \toprule[1.5pt]
    \end{tabular}
  \end{center}
  \vspace{-0.7cm}
\end{table*}
}%
\section{Simulations and Results}\label{simulation}
In this section, we use the proposed and developed NAS algorithm to select the proper CNN for different signal sets which are representing the features of wireless channel and noise in a variety of physical environments. Here, three signal sets are prepared, and the proposed NAS algorithm offers three CNNs with the highest achieved $P_c$ in all generated episodes for each dataset. These signal sets are generated according to (\ref{eq2}) in two classes of vacant and active sensed frequency channels. \par
{\it 1)} In the first dataset, we consider a conventional model in the literature with circularly symmetric complex white Gaussian (CSCWG) noise, $w[n]\sim\mathcal{CN}(0,\sigma^2_w)$. Accordingly, 20K signals are generated for $H_0$. For the signals of $H_1$ hypothesis, the fading channel is modeled as flat fading where $h[n]=d_0\delta[n]$, and  $d_0\sim\mathcal{CN}(0,1)$. PU's signal is distributed by $s[n]\sim\mathcal{CN}(0,\sigma^2_s)$. SNR of signals are defined as $\mathsf{SNR}=\sigma^2_s/\sigma^2_w$ and picked from $\{-20, -18, ..., 18\}$ dB. The total number of signals with $N=100$ samples in each signal is 40K in this dataset.\par
{\it 2)} The second dataset also follows (\ref{eq2}) while impulsive noise distributed by complex isometric S$\alpha$S ($w[n] \sim S\alpha S(\alpha,\mu, \gamma)$). Here, $\alpha$, $\mu$, and $\gamma$ are exponent, location, and dispersion factors of distribution \cite{NOLAN_1general,Kang2010ACO}. In \cite{Lunden_SS_imp}, the empirical data is finely consistent  with S$\alpha$S model with $\alpha$ close to $1.25$. Hence, we use $w[n]\sim S\alpha S(1.25,0, 1)$.
The signal of PU is OFDM with symbols of QPSK and 64 subcarriers where it is equal to the points of the fast Fourier transform (FFT) $N_{fft}$. Following IEEE 802.11 standard \cite{goldsmith2005wireless}, 12 subcarriers on two sides of OFDM symbol are as guards, and four subcarriers are for pilot symbols. A cyclic prefix with the length of 16 is added to create an OFDM symbol with the length of 80 points. Each QPSK symbol has a time symbol of $T_s=50$ ns which results in an OFDM symbol with a time length of $T_{ofdm}=80T_{s}=4$ \textmu s. In contrast to the first dataset, a frequency selective channel with a tapped-delay impulse response of $h[n]=\sum_i d_i\delta(n-n_i)$ is used where delays, $n_i$, and amplitudes  of taps, $d_i$, come from the extended pedestrian A (EPA) model with the maximum time delay of $410$ ns \cite{EPA}. The sensing time in this case is set to $T_{sen}=8$ \textmu s, which is proportionate to the time of two OFDM symbols. We assume the sampling time is equal to the symbol time  $T_s=50$ ns$=1/f_s$ which results in $160$ samples in the sensing time. For this impulsive noise model, the generalized signal-to-noise ratio (GSNR) is defined as $\mathsf{GSNR}=P/4\gamma$ where $\sqrt{P}$ is the amplitude of PU's signal. Signals from $H_1$ have a $\mathsf{GSNR} \in \{-5,-4,...,24\}$ dB with a total number of 15K signals while number of signals from $H_0$ is 15K, which collectively form the dataset.\par
{\it 3)} The third dataset, in comparison with the previous dataset, has similar noise, signal, channel model, and number of signals. However, it benefits from a quadrupled sensing time, $T_{sen}=32$ \textmu s with the duration of eight OFDM symbols and $N=640$ samples. The amounts of $\mathsf{GSNR}$  for signals in class of $H_1$ is $\{-10,-9,...,20\}$ dB. \par
These three different datasets are employed for the proposed NAS algorithm in order to specify the reward for the networks. Three types of convolutional, max-pooling, and GAP layers are included. For convolutional layers, the number of filters are selected from $n_f^{<l>}\in\{8,16,32,64\}$, and the size of filters are picked from  $s_f^{<l>}\in\{3,5\}$. For max-pooling layers, two options for the size of pooling is assumed, $s_p^{<l>}\in\{2,4\}$. The maximum number of $11$ options for actions in each state ($8$ different convolutional layers, $2$ different pooling layers, and GAP layer) is possible. The length of networks is limited to $L=8$, and the minimum possible length of signal is set to $N_{trsh}=8$.  Since the agent is not capable of returning to a previous state, no loops exist in the MDP. The number of networks described by this MDP is approximately between $8^8$ and $11^8$. Enlarging the action space offers more options to the agent for gaining a higher accuracy at the cost of an extreme growth of the search space. Using the proposed algorithm, the agent  attempts to find the network with the highest achieving reward. The output of a 10-fold validation of networks  on each dataset is regarded as $P_c$ and returned to the agent as a reward for updating action-value functions.
 Through the generated episodes,  $\epsilon$ is decreased from 1 (a full exploratory policy) to 0 (the greedy policy). Table \ref{tab3} shows the best network with the highest achieved reward for each dataset. Also, information of datasets, the required run time of algorithm in GPU days (number of GPUs multiplied by days), and the mean required time of training per epoch for each dataset is reported  in this table. As seen, the mean training time for more complex architectures for Dataset 3 is longer. This indicates that for a short-term updating such as fine-tunning or weight updating of network with new signals \cite{peng}, it takes about 2.42 seconds for each epoch. Also, for long-term update of the architecture, use of the previous $Q([\textbf{s}_l,\textbf{a}_l])$ as the initial value for NAS algorithm can significantly reduce the run time. For the implementation of this method, Python and TensorFlow are used with NVIDIA P6000 GPU. Interestingly, in all of the three suggested architectures for CNN, max-pooling  layer is not employed, and for more complicated datasets with larger signal samples, a more complex network is suggested.\par
{\renewcommand{\arraystretch}{1.1}
\begin{table}[t!]
\fontsize{9pt}{10pt}\selectfont 
  \begin{center}
    \caption{{\fontsize{9pt}{9pt}\selectfont Accuracy of different networks and  proposed NAS-CNNs for different datasets.} \label{tab4}}

    \begin{tabular}{
 >{\centering}m{0.11\textwidth}
 >{\centering}m{0.09\textwidth}
  >{\centering}m{0.09\textwidth}
    >{\centering\arraybackslash}m{0.09\textwidth}} 
  \toprule[1.5pt]
&\multicolumn{3}{c}{\textbf{Accuracy} $P_c (\%)$ }
   \\
   \cmidrule[1.5pt](l{0pt}r{0pt}){2-4}
\textbf{Detector}&\textbf{Dataset 1} &\textbf{Dataset 2}& \textbf{Dataset 3}\\
\toprule[1.5pt]
CM-CNN  &  77.64 & 49.96 & 49.26\\
      \hline
      CNN    &  77.52 & 75.20 & 65.94\\
     \hline
     FLOM-CNN    & 79.12   &63.36 & 56.97\\
      \hline
     SAE   &   69.89 & 49.88& 49.95\\
      \hline
       LSTM    &  79.44  & 76.28 & 57.39\\
      \hline
       NAS-CNN    & 79.61   & 78.29  &75.25\\

      \toprule[1.5pt]
    \end{tabular}
  \end{center}
  \vspace{-0.7cm}
\end{table}
}%
Table \ref{tab4} contains the accuracy $P_c$ of  CNNs architecture suggested by the NAS algorithm (NAS-CNN) for the corresponding dataset. These figures are computed by use of 10-fold validation with 15  epochs for Dataset 1, and 20 epochs for Dataset 2 and 3.  Moreover, for the sake of comprehensive comparisons, accuracy of the proposed networks with different types of layers in the literature such as covariance matrix-aware CNN (CM-CNN) \cite{CLiu}, CNN \cite{mehrabian_new}, fractional lower-order moments CNN (FLOM-CNN) \cite{flom-cnn}, SAE \cite{QCheng}, and LSTM \cite{LSTM} for these datasets are presented in this table. In the following, in addition to the mentioned data-driven networks with various architectures and types of layers, different model-based detectors especially detectors suitable for impulsive noise in the literature such as FLOM \cite{XZhu} and Cauchy detector (CD) \cite{Kang2010ACO} are employed for comparison. As seen, NAS-CNNs achieve higher accuracy in comparison with other detectors. Except for Dataset 3, LSTM has the best performance after the suggested NAS-CNNs. CM-CNN's performance is not suitable for S$\alpha$S since the amount of the inputted covariance matrix to the network and second-order moments of the signal diverge under impulsive noise \cite{samoradnitsky2017stable}.

\begin{figure}[t!]
\centering
		\includegraphics[width=3.5in]{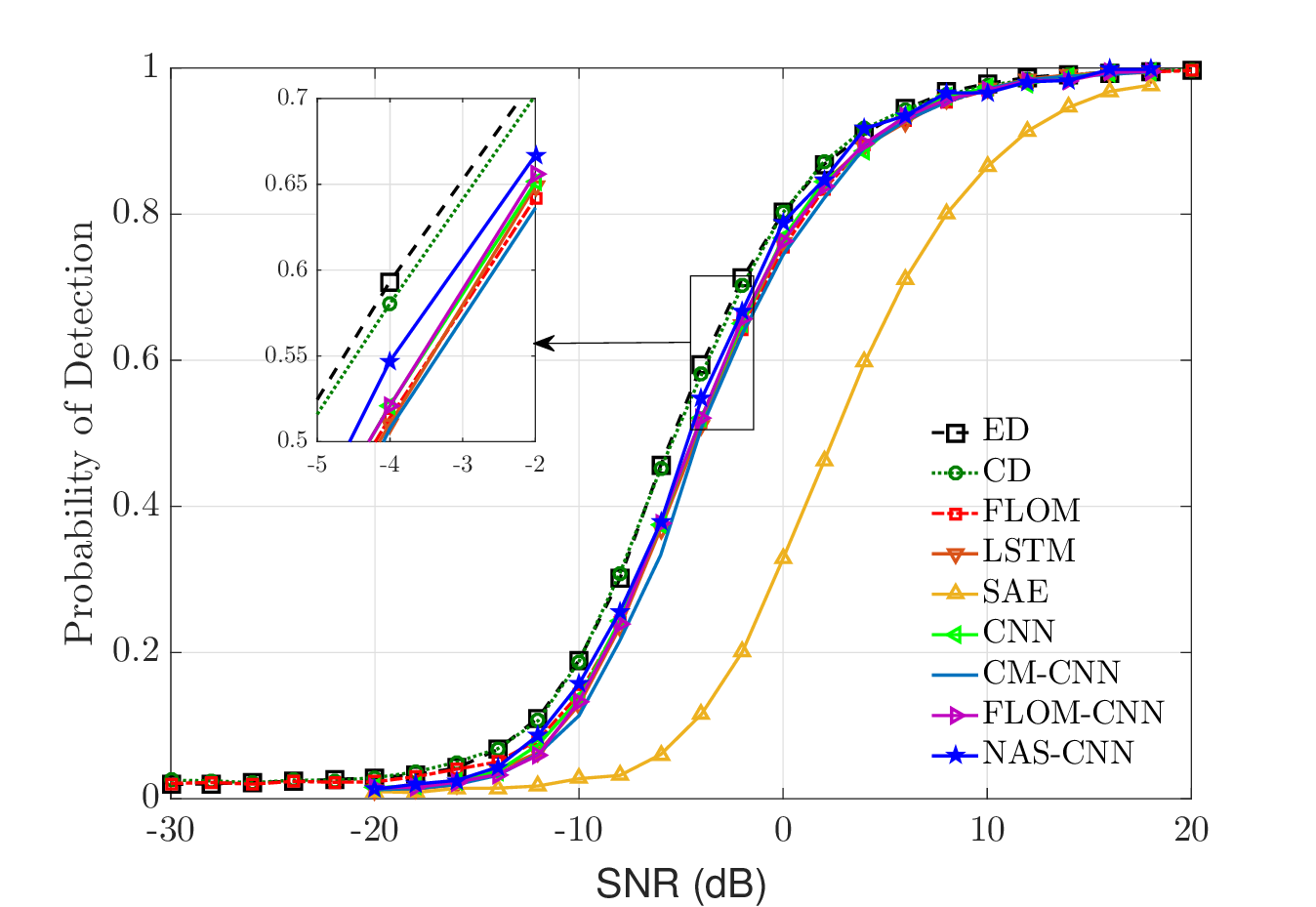}
	\caption{{\fontsize{9pt}{9pt}\selectfont $P_d$ versus SNR for detectors under Gaussian signal and noise.}{\label{fig44}} }
	\vspace{-0.5cm}
\end{figure}

The first dataset models general signals with a convenient noise and channel model. ED is the optimum solution in the case of both Gaussian noise and Gaussian signal models \cite{Axell2011Optimal, kay1993fundamentals}. Fig. \ref{fig44} with the curve of detection probability versus SNR of ED confirms this fact. Performance of other model-based detectors including FLOM and CD are  also shown in this figure. In fact, except for SAE, detection performance of  all networks is close to each other, and they follow  curves of ED and CD with a slight gap. Although ED has the highest detection performance and benefits from a simple implementation, its performance degrades dramatically under noise uncertainty \cite{Taherpour} and general noise distributions \cite{mehrabian_new}.

\subsection{Receiver Operating Curve }
\begin{figure}[t]
\centering
		\includegraphics[width=3.3in]{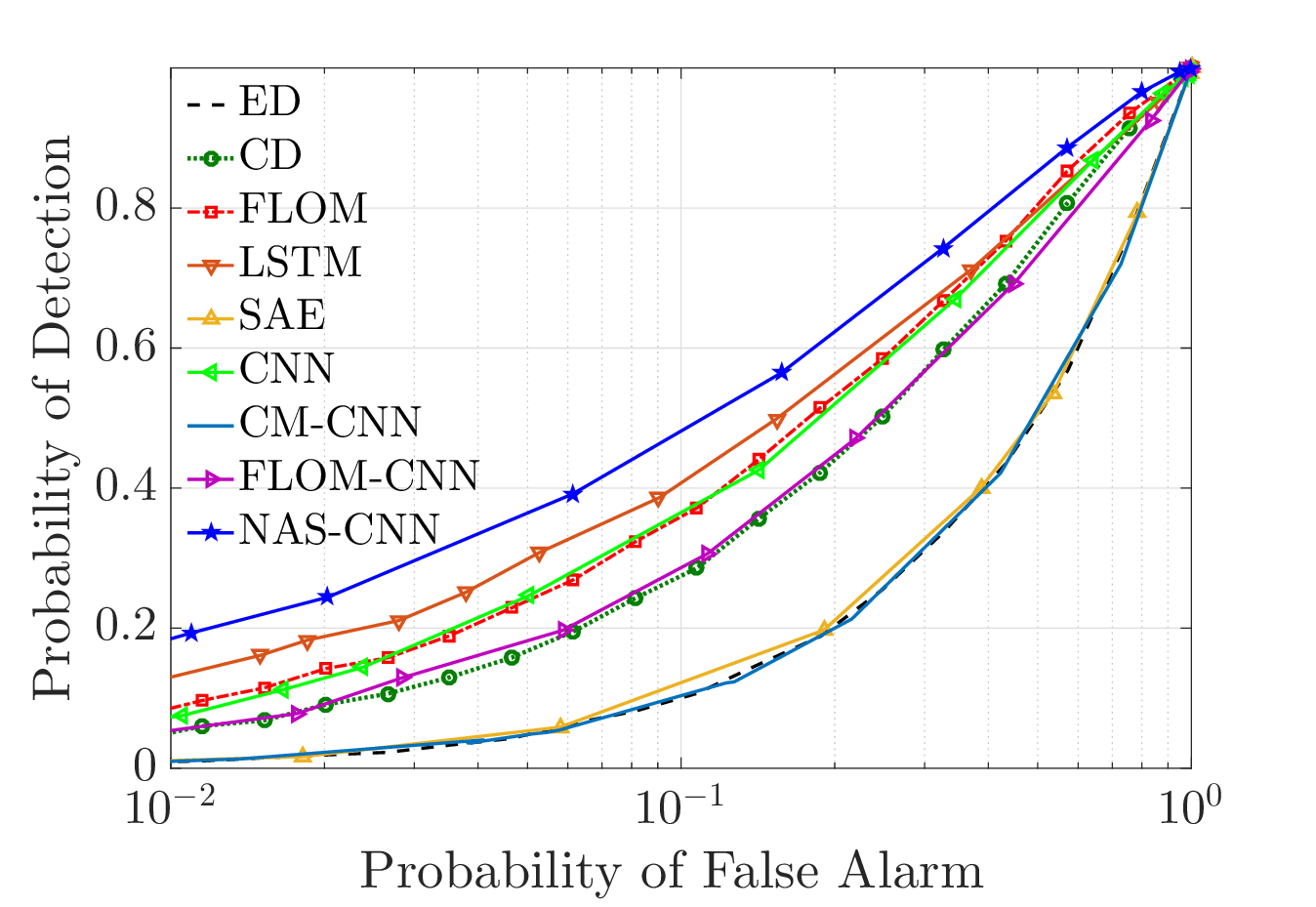}
	\caption{{\fontsize{9pt}{9pt}\selectfont ROC of detectors for Dataset 2 and  test signals with $\mathsf{GSNR}=5$ dB.}
	{\label{fig5}} } 
	 \vspace{-0.7cm}
\end{figure}

\begin{figure}[t]
\includegraphics[width=3.3in]{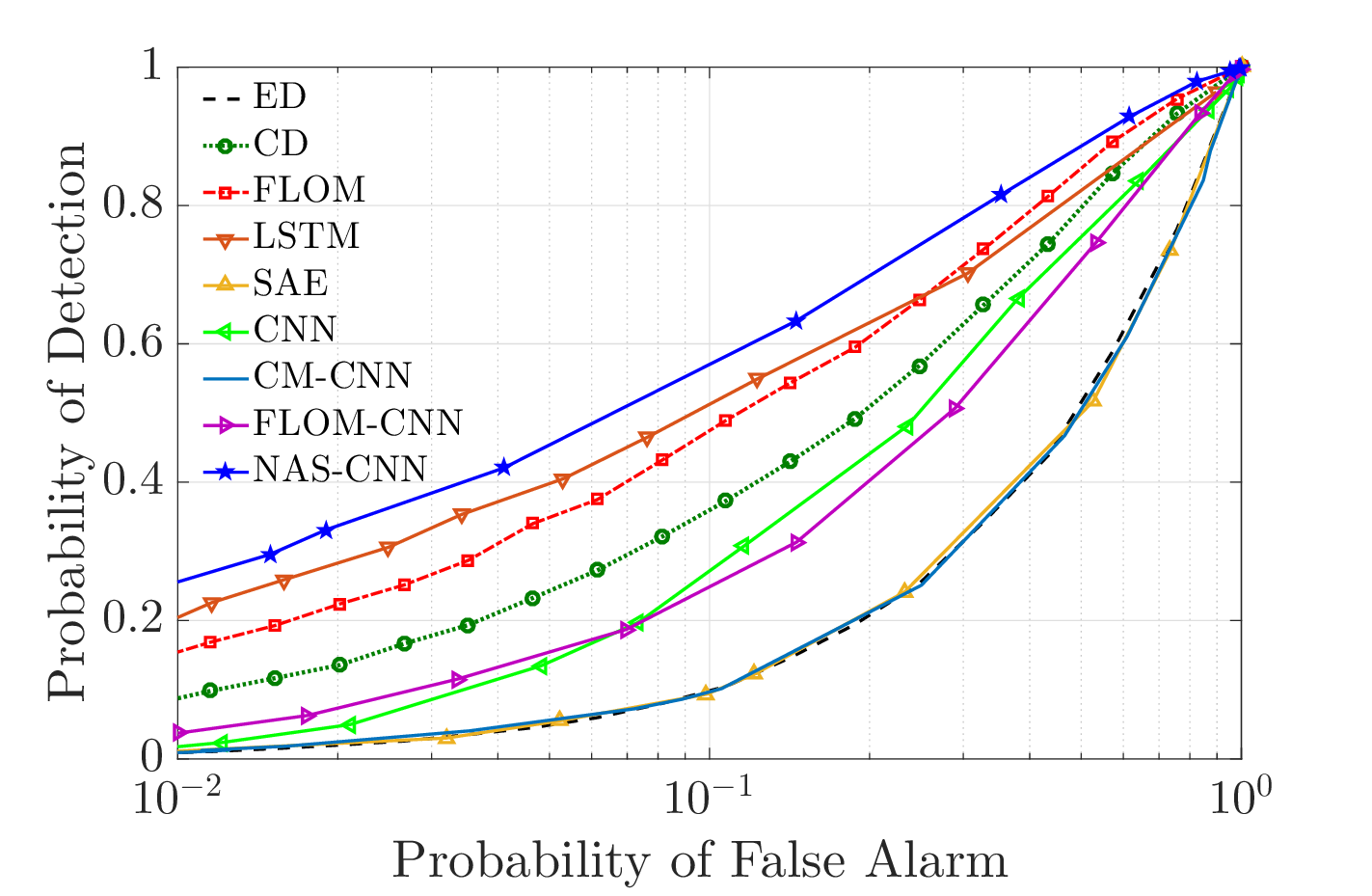}
	\caption{{\fontsize{9pt}{9pt}\selectfont ROC of detectors for Dataset 3   and test signals with $\mathsf{GSNR}=3$ dB.}
	{\label{fig6}} }
 \vspace{-0.7cm}
\end{figure}

Here, we scrutinize the performance of the proposed detectors by NAS-CNN and the above-mentioned detectors in the literature for  OFDM signal of IEEE 802.11 with EPA  wireless channel and a general noise model of isometric complex S$\alpha$S. We used probability of detection versus probability of false alarm, called receiver operating curve (ROC), to examine the performance of the proposed NAS-CNN and others. Datasets 2 and 3 with short and long sensing times are used for training the DNN-based detectors and also adjusting the detection threshold for the rest of the simulations. In Fig. \ref{fig5} for finding probability statistics, parameters of test signals are generated based on  parameters used in Dataset 2 but with $\mathsf{GSNR}=5$ dB. As seen, in this case, the NAS-CNN offers the best detection probability, and in the second rank, LSTM has the greatest detection probability. At the other extreme, ED, CM-CNN, and SAE have the lowest detection probability. It was anticipated for ED and CM-CNN since energy and covariance matrix diverge under impulsive noise. The same result is observed in  Fig. \ref{fig6} where Dataset 3 with 8 OFDM symbols   is used for training. The generated test signals have a $\mathsf{GSNR}=3$ dB. Detection probability for NAS-CNN architecture surpasses that of other detectors. Accordingly, we conclude that for both these scenarios, CNN architectures suggested by the developed NAS method have a superior ROC in comparison to others.
\subsection{Probability of Detection Versus GSNR }
\begin{figure}[t]
		\includegraphics[width=3.5in]{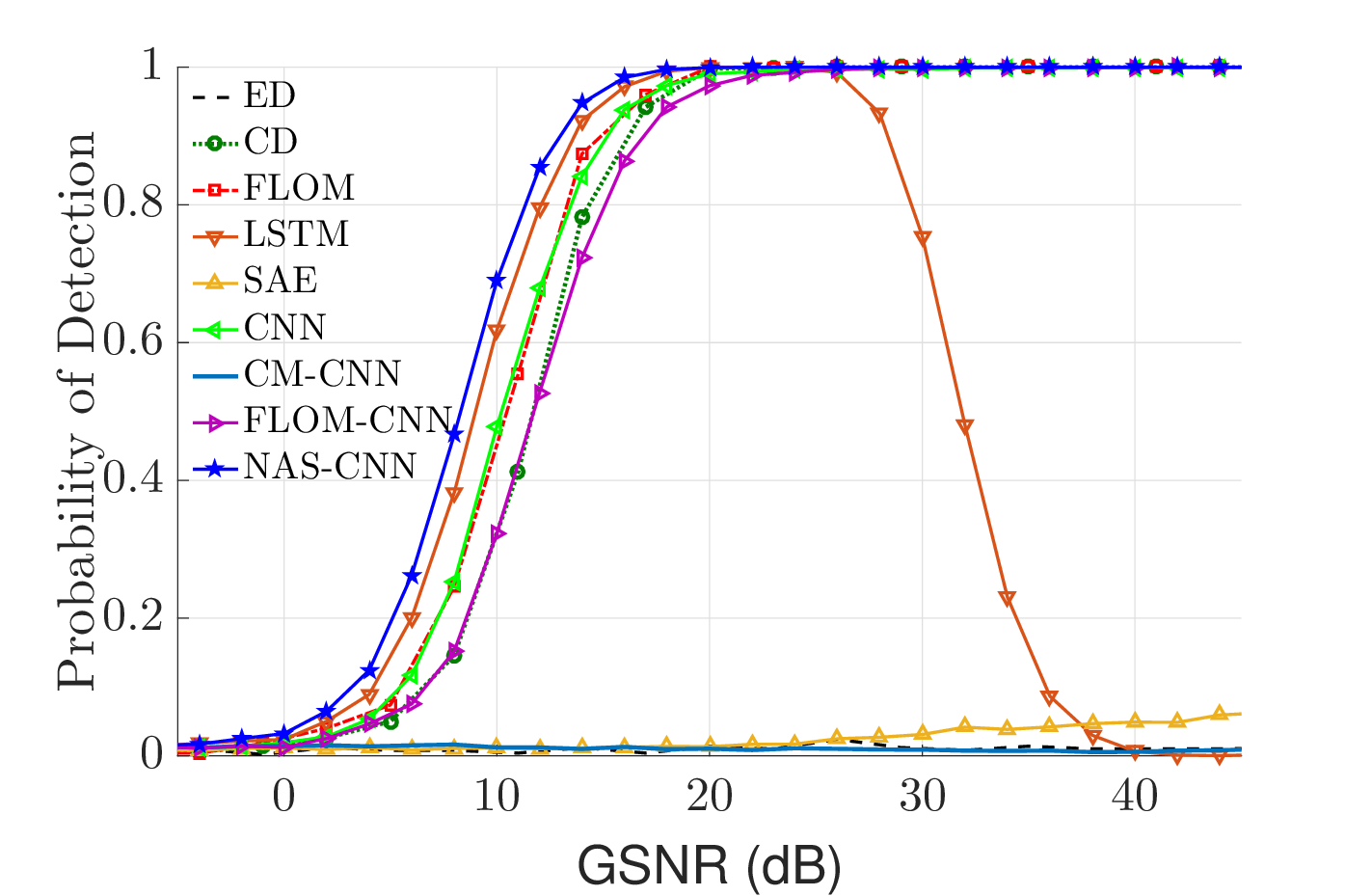}
	\caption{{\fontsize{9pt}{9pt}\selectfont $P_d$ versus GSNR for detectors trained by Dataset 2 with $T_{sen}=8$ \textmu s.}
	{\label{fig7}} }
	  \vspace{-0.5cm}
\end{figure}

For the following simulations, the detection threshold for detectors is adjusted to guarantee $P_{fa}<p_{fa}=0.01$, and then, the detection probability of these methods, $P_d$, is found for different amounts of GSNR. Both Dataset 2 and 3 are used for training the networks proposed by NAS (architectures in rows 2 and 3 of Table \ref{tab3}) and other DNN-based detectors. For test signals, the same set of parameters are used; however, GSNR has a broader range, and test set includes signals with distinctive GSNR in comparison to those in training datasets. As observed in Fig. \ref{fig7}, $P_d$ is obtained for a broad range of GSNR to investigate the performance of networks under test signals with a variety of signal strength which also are not used in the training signal set.
In this figure, the sensing time is short $T^s_{sen}=2T_{ofdm}=8$ \textmu s, and complex isometric  S$\alpha$S noise with an exponent parameter of $\alpha=1.25$ and a dispersion parameter of $\gamma=1$ with EPA channel is considered. Fig. \ref{fig7} shows the superiority of the proposed NAS architecture for this signal setup. LSTM network also performs properly for $\mathsf{GSNR}<26$ dB, and its performance severely degrades for  GSNRs beyond that. In contrast, networks based on convolutional layers including NAS-CNN perform robustly in the entire range of  GSNR.\par
\begin{figure}
		\includegraphics[width=3.5in]{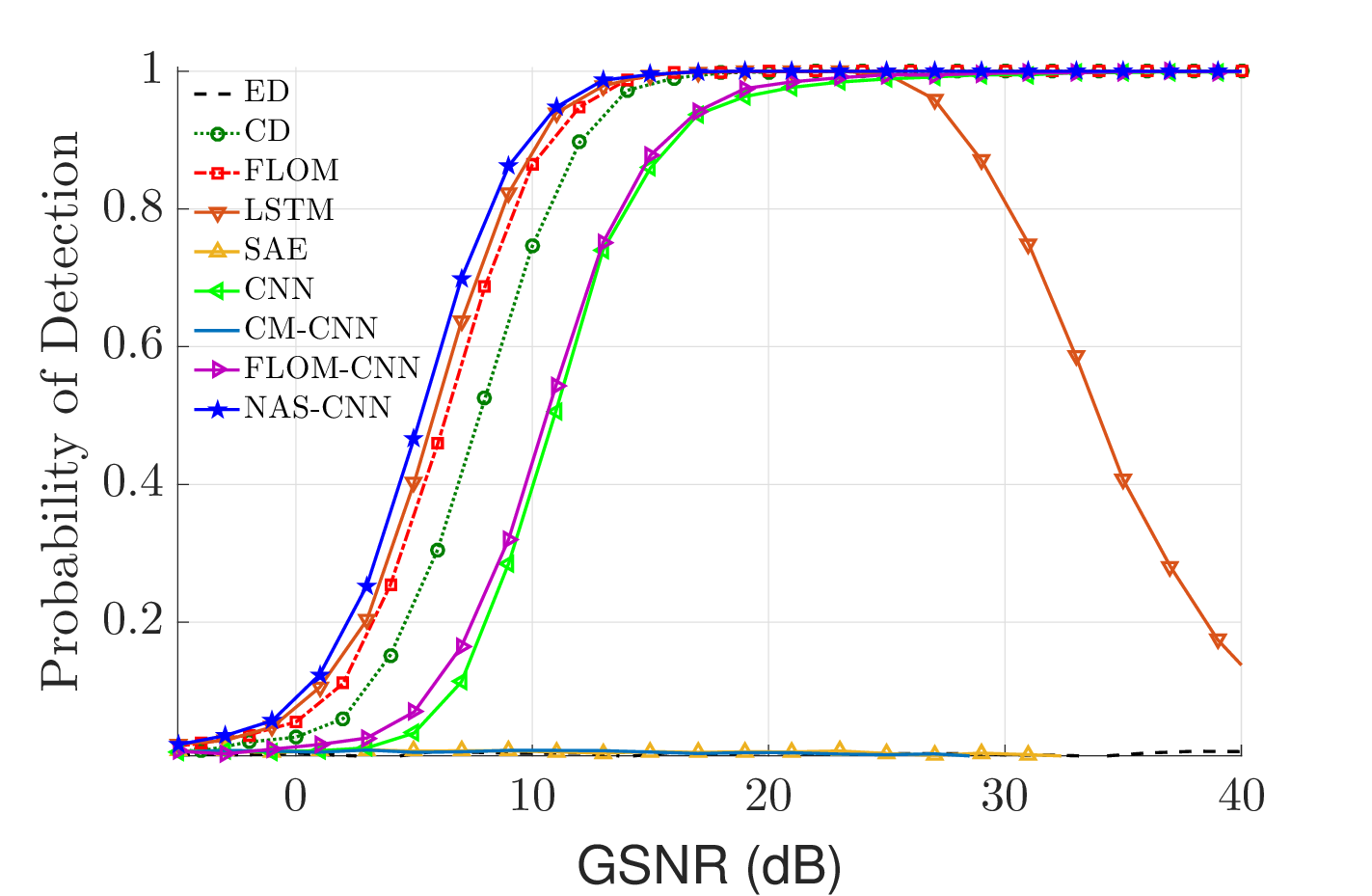}
	\caption{{\fontsize{9pt}{9pt}\selectfont $P_d$ versus GSNR for detectors trained by Dataset 3 with $T_{sen}=32$ \textmu s.}
	{\label{fig8}} }
  \vspace{-0.7cm}
\end{figure}

We proceed with the similar simulation for the Dataset 3  which corresponds to a sensing time equal to $T^l_{sen}=8T_{ofdm}=32$ \textmu s. Fig. \ref{fig8} contains the detection probability of these detectors. The same result can be observed in this figure where NAS-CNN in row 3 of Table \ref{tab3}, LSTM, FLOM, and CD offer the greatest probability of detection, respectively. LSTM's curve similar to the previous case starts to descend for $\mathsf{GSNR}>25$ dB. This suggests that the network with LSTM cells, in contrast to CNNs, is more prone to become overfitted to signals with GSNRs in the trained dataset leading to its performance decline for detection of test signals with large GSNRs.\par
It was anticipated that increasing the sensing time results in a higher probability of detection, and this result can be easily concluded  by comparing Fig. \ref{fig7} and \ref{fig8}. As an example, in Fig. \ref{fig7} for $\mathsf{GSNR}=10$ dB with $T^{s}_{sen}=8$ \textmu s, the probability of detection obtained by NAS-CNN is about $P_d=0.7$ while NAS-CNN designed for a sensing time of $T^{l}_{sen}=32$ \textmu s achieved $P_d=0.9$. Therefore, we use these results for different sensing times in order to improve the whole performance of the  SU by employing the RL method explained in Section \ref{RLforSensing}.\par

\subsection{Discrepancy of Parameters and Complexity of Methods}
\begin{figure}[t]
	\includegraphics[width=3.5in]{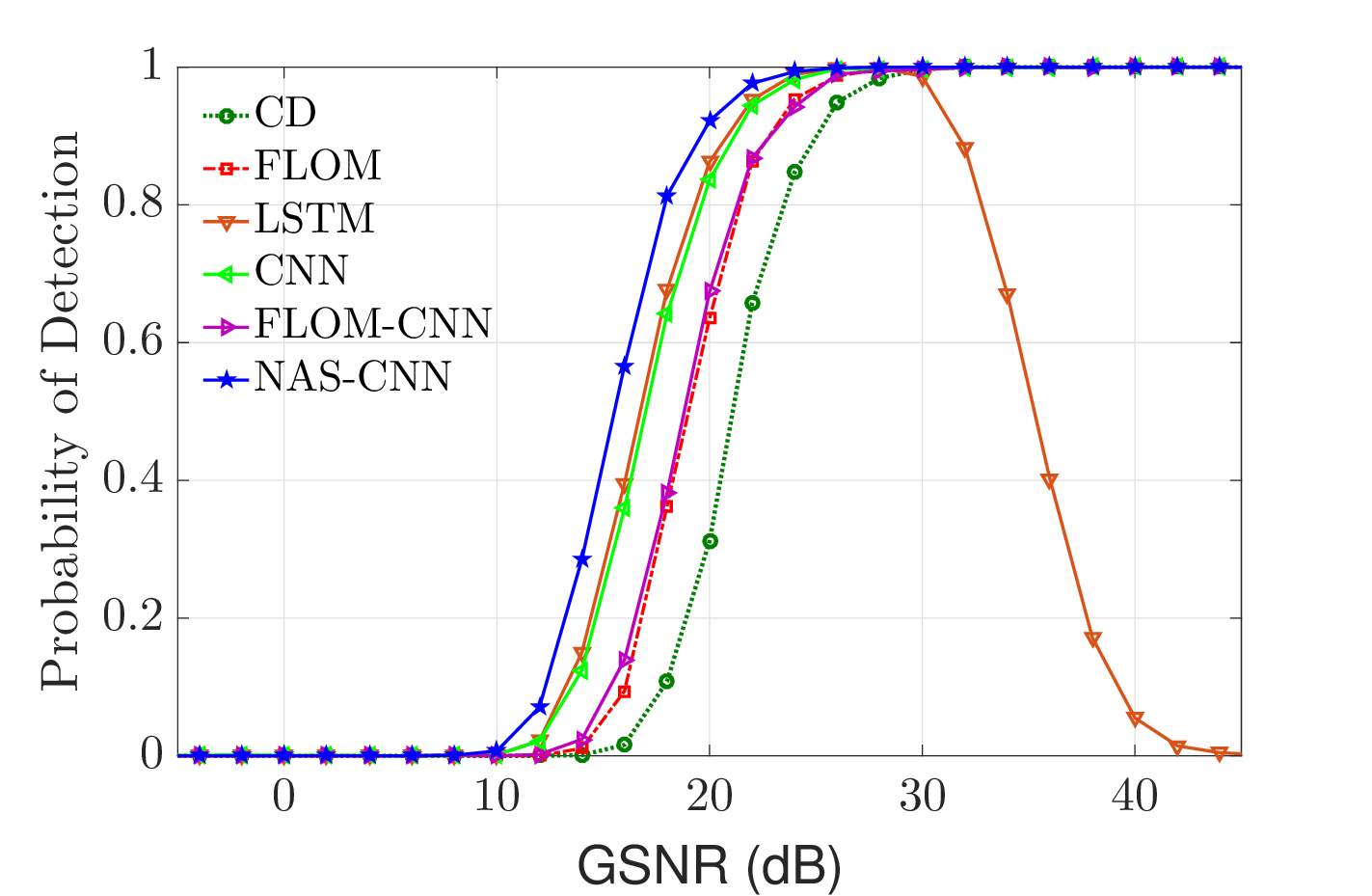}	
	\caption{{\fontsize{9pt}{9pt}\selectfont $P_d$ versus GSNR for detectors trained by Dataset 2 and tested with $\gamma^{test}=0.5$. 
	}
	{\label{fig9}} }
\vspace{-0.5cm}
\end{figure}

\begin{figure}
		\includegraphics[width=3.5in]{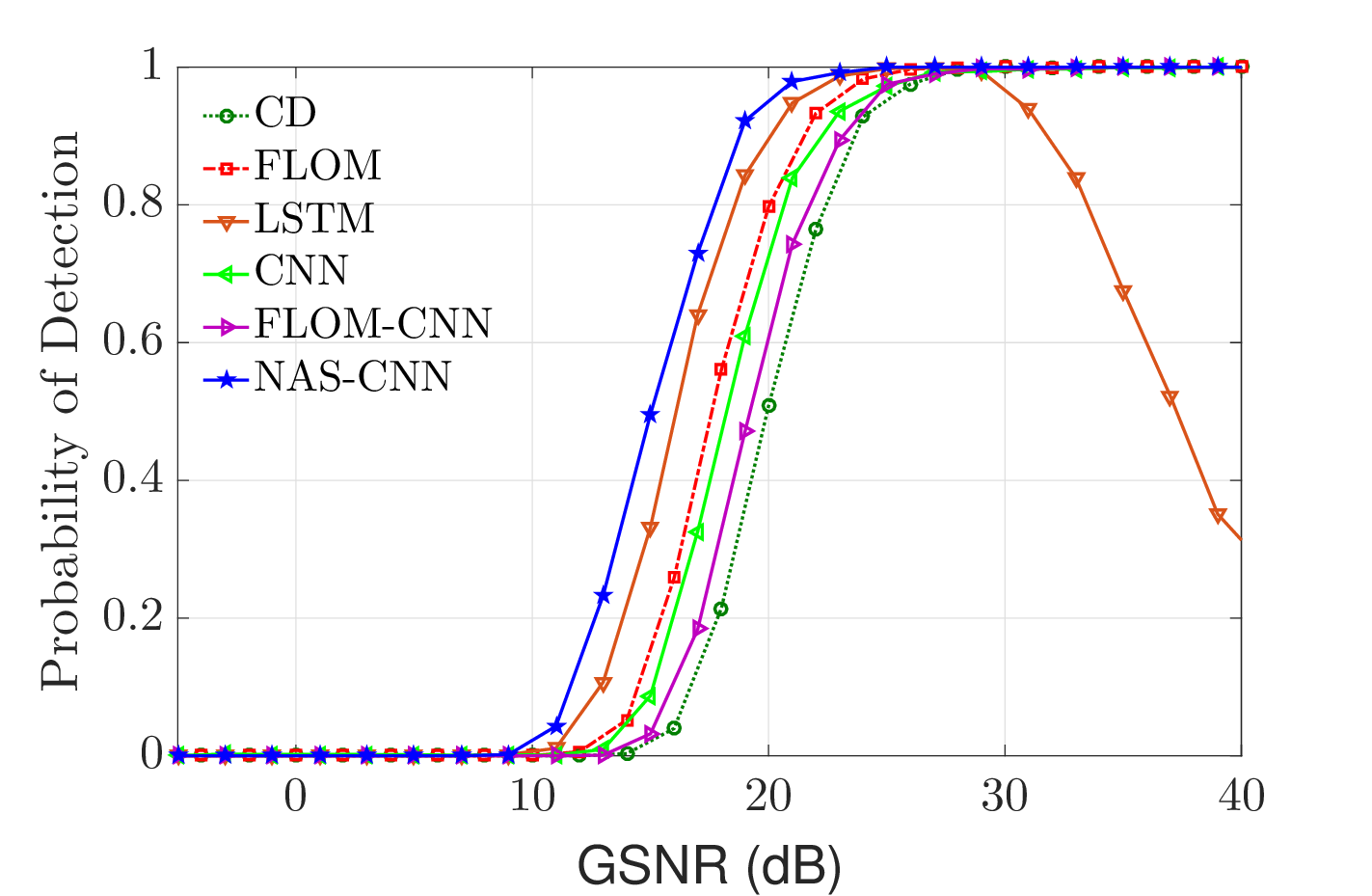}	
	\caption{{\fontsize{9pt}{9pt}\selectfont $P_d$ versus GSNR for detectors trained by Dataset 3 and tested with $\gamma^{test}=0.5$. 
	}
	{\label{fig10}} }

 \vspace{-0.7cm}
\end{figure}

In the following, networks are trained using parameters of Dataset 2 and 3. Then, signals generated with different parameters from the used parameters in these datasets are employed to  evaluate the generalization ability and robustness of networks under mismatch of parameters. After training process with $w[n]\sim S\alpha S(\alpha=1.25,\mu=0,\gamma=1)$, test signals are produced with the same noise parameters and $50\%$ difference in dispersion parameter where $\gamma^{test}=0.5$. 
 Fig. \ref{fig9} presents $P_d$ for various GSNRs where the sensing time is $T^{s}_{sen}=8$ \textmu s. The best detection performance belongs to NAS-CNN followed by LSTM, CNN, FLOM-CNN, FLOM, and CD while we excluded other detectors since we observed their weak performance  in the previous simulations. LSTM and CNN in this setup have a close performance to each other. By considering  Fig. \ref{fig7} and \ref{fig9}, performance of all detectors are degraded due to $50\%$ mismatch of parameters in dispersion parameter $\gamma$. The same degradation  is observed by comparing  Fig. \ref{fig8} with curves of Fig. \ref{fig10} which shows $P_d$ under $\gamma^{test}=0.5$ and with $T^l_{sen}=32$ \textmu s. Similarly, the proposed NAS-CNN with a longer input signal attains the top performance regarding all its counterparts in this figure.
Accordingly, 
NAS-CNNs obtain the highest detection probability  in all cases of OFDM signals with impulsive noise, and for the case of Gaussian noise and Gaussian signal,  performance of NAS-CNN is close to that of the model-based ED, which is regarded as the optimum detector in this case \cite{Axell2011Optimal,kay1993fundamentals}.\par
Similar to \cite{QCheng,mehrabian_new} for comprehensive comparisons, we present the number of the required real multiplications (RRMs)  and the number of weights in networks  for online detection in these three datasets to investigate the  complexity of these methods. Table \ref{tab55}  includes the number of RRM (\#RMM) and weights (\#W) in millions (M) and thousands (K), respectively, for CNN, CM-CNN, FLOM-CNN, SAE, and NAS-CNN. 
In terms of the least RRM and weights, NAS-CNN for Dataset 1 is in the third and first place, respectively. Its number of RRM is also  in the same order of SAE and FLOM-CNN. After FLOM-CNN, NAS-CNN has the least number of parameters for Dataset 2 and 3 in comparison with other detectors in Table \ref{tab55}. However, due to multiple convolutional layers, NAS-CNN should compute a great number of RRM where this number increases by a longer sensing time in Dataset 3. Thus, NAS-CNN has the highest number of RRM in online detection of signals in Dataset 3.
 \ We can punish the agent by adding negative rewards  to reward function proportionate to the complexity of the proposed network. This implies a trade-off between complexity and accuracy of a network, and one can adjust the reward to force the agent to prefer more accurate networks or less complex ones. 
 \subsection{Sensing Time Selection}
\begin{figure}[t!]
		\includegraphics[width=3.5in]{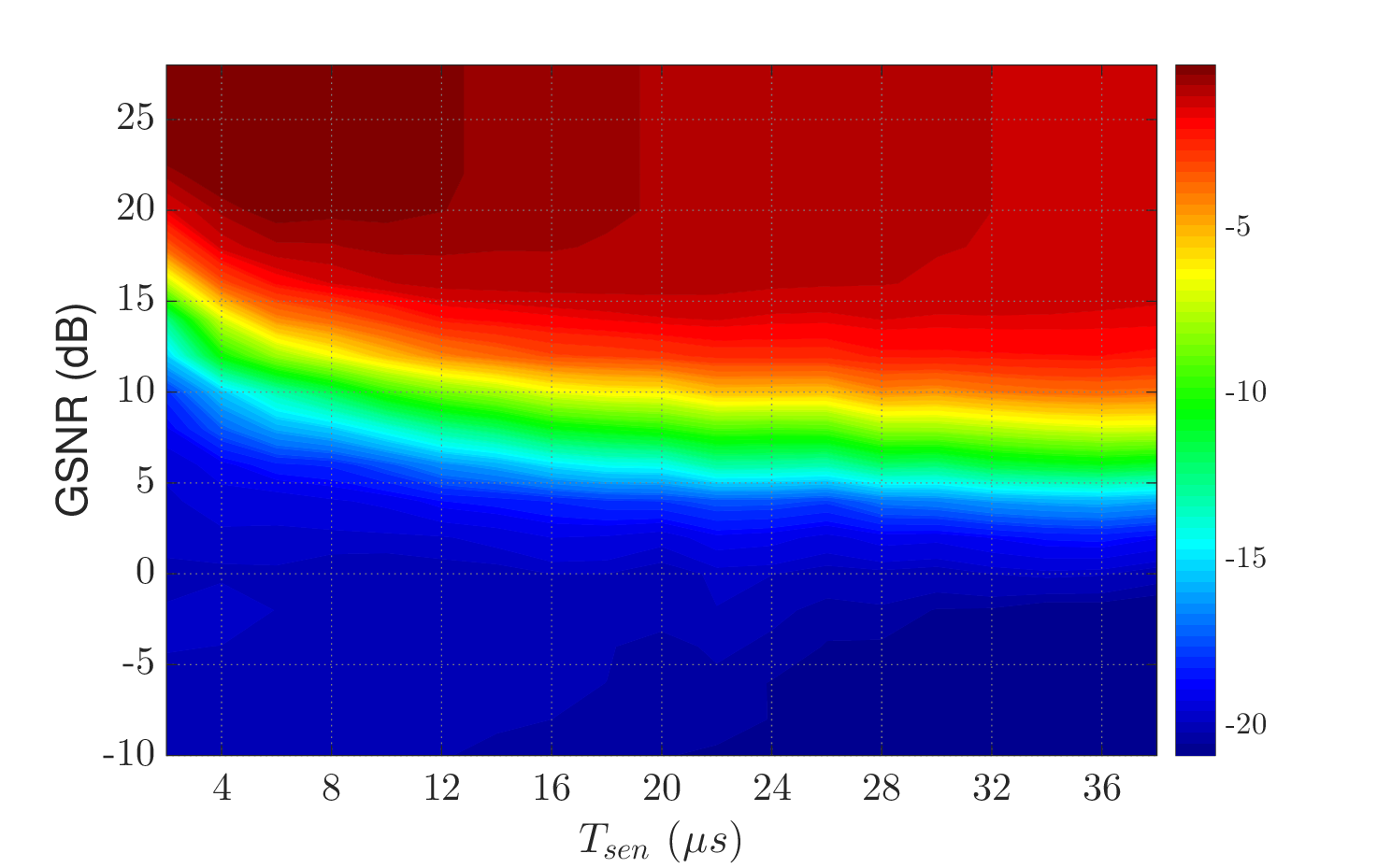}
\centering	
	\caption{{\fontsize{9pt}{9pt}\selectfont Average reward gained by FLOM detectors with various sensing times for different GSNRs.}
	{\label{fig_R_TS_SNR}} }
	\vspace{-0.5cm}
\end{figure}

\begin{figure}[t!]
		\includegraphics[width=3.5in]{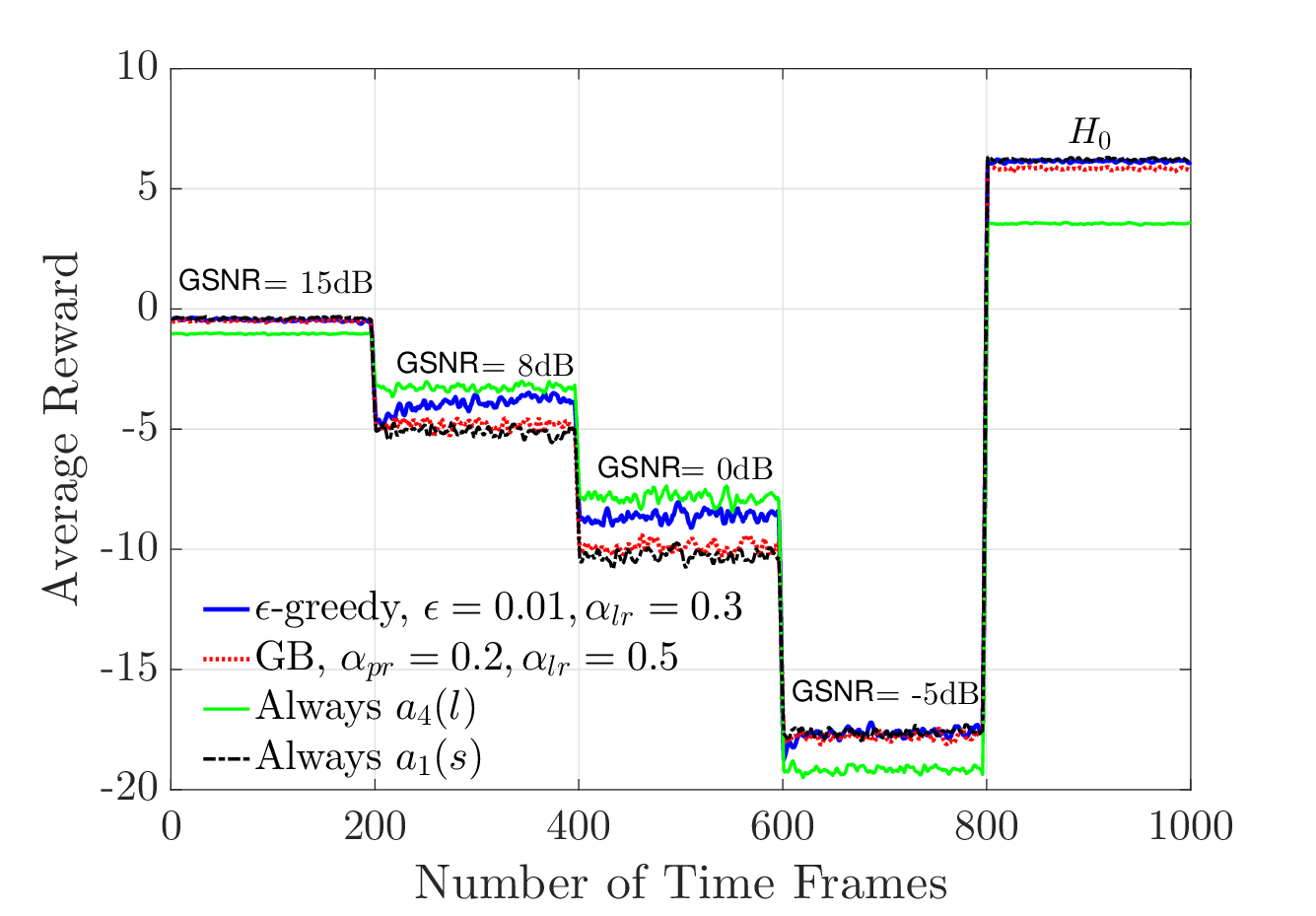}
\centering	
	\caption{{\fontsize{9pt}{9pt}\selectfont Smoothed average reward of policies with NAS-CNN detectors in different time frames.}
	{\label{fig11}} }
	\vspace{-0.7cm}
\end{figure}

 In Sections \ref{RLforSensing}, we discussed the proposed method of selecting a sensing time from a pool of discrete options. 
 Here, we use the proposed RL method for selecting the sensing time with FLOM  detector \cite{XZhu} with various lengths and also with NAS-CNN of the previous sections designed for sensing time of 2 and 8 OFDM symbols. 
{\renewcommand{\arraystretch}{1.1}
\begin{table*}[t!]
\fontsize{9pt}{10pt}\selectfont 
  \begin{center}
    \caption{{\fontsize{9pt}{9pt}\selectfont Number of required real multiplications and weights for various methods of detection.} \label{tab55}}

    \begin{tabular}{
 >{\centering}m{0.16\textwidth}
 >{\centering}m{0.18\textwidth}
  >{\centering}m{0.18\textwidth}
    >{\centering\arraybackslash}m{0.18\textwidth}} 
  \toprule[1.5pt]
\multirow{2.3}{*}{\textbf{Detector}}&\multicolumn{3}{c}{\#RRM (M) - \#W (K) }
   \\
   \cmidrule[1.5pt](l{0pt}r{0pt}){2-4}
&\textbf{Dataset 1} &\textbf{Dataset 2}& \textbf{Dataset 3}\\
\toprule[1.5pt]
 CNN  &  4.173 - 509.4 & 6.693 - 771.6 & 26.722 - 2737.7\\
      \hline
      SAE    &  0.025 - 25.3 & 0.037 - 37.3 &  0.133 - 133.3\\
     \hline
     FLOM-CNN    & 0.028 - 0.8   & 0.028 - 0.8 & 0.028 - 0.8\\
      \hline
     CM-CNN  &   0.826 - 127.6 & 0.826 - 127.6 & 0.826 - 127.6\\
      \hline

       NAS-CNN    & 0.038 - 0.5   & 4.854 - 30.6 & 27.075 - 42.7\\

      \toprule[1.5pt]
    \end{tabular}
  \end{center}
  \vspace{-0.2cm}
\end{table*}
}%
 We set the components of the aggregate reward including throughput, interference, and complexity as $\lambda_1T_{0,0}^a=\lambda_1R_{SU}(T_{f}-T^a_{sen})=(T_{f}-T^a_{sen})/10$,
 $\lambda_2\xi=20$,
  and $\lambda_3C^{a}_{i,j}=\lambda_3T^a_{sen}=T^a_{sen}/32$ where $T_{f} =20T_{ofdm}=80$ \textmu s.
 Fig. \ref{fig_R_TS_SNR} shows the average reward of a detector for various GSNRs and $T_{sen}$ for FLOM detector. As seen, it is clear that for middle GSNRs, longer sensing times (due to less misdetection) and for small and very large GSNRs shorter sensing times (due to less complexity) achieve higher average reward, which is consistent with the findings reported in Table \ref{tab2}.
To examine the performance of policies in (\ref{eq11}) and (\ref{eq13}), we use two NAS-CNN architectures in Table \ref{tab3} for OFDM signal. For both policies, the selected action $a$ belongs to an action space with two actions $a\in\mathcal{A}_2=\{a_1=s,a_4=l\}$ where $T^a_{sen}\in\{T^{a_1}_{sen}=8, T^{a_4}_{sen}=32\}$ \textmu s, and the two actions are called short ($s$) and long ($l$) sensing times.
1000 consecutive generated time frames, shown in Fig. \ref{fig11}, are divided into five sections with 200 time frames where in each of these five sections the channel condition $(\mathsf{GSNR},H_i)$ is simulated based on  $(15 \mathrm{dB},H_1)$, $(8 \mathrm{dB},H_1)$, $(0 \mathrm{dB},H_1)$, $(-5 \mathrm{dB},H_1)$, and $(-,H_0)$, respectively.
Clearly, this simulation shows a non-stationary environment in which $H_i$ and GSNR of the channel change in different sections. 
 Four policies in this scenario are used for comparison. Both of policies are $\epsilon$-greedy and GB in (\ref{eq11}) and (\ref{eq13}) where parameters of $\epsilon$, $\alpha_{lr}$ and $\alpha_{pr}$ for these policies are opted with  grid search. Additionally, for policies with  always long and always short actions, which are commonly employed in the literature, 
 just one NAS-CNN with long (always long) or short (always short) sensing time is used for the detection in all time frames of this simulation. Fig. \ref{fig11} shows the average reward curve of four policies through all 1000 time frames.\par
\begin{figure}[t!]
		\includegraphics[width=3.5in]{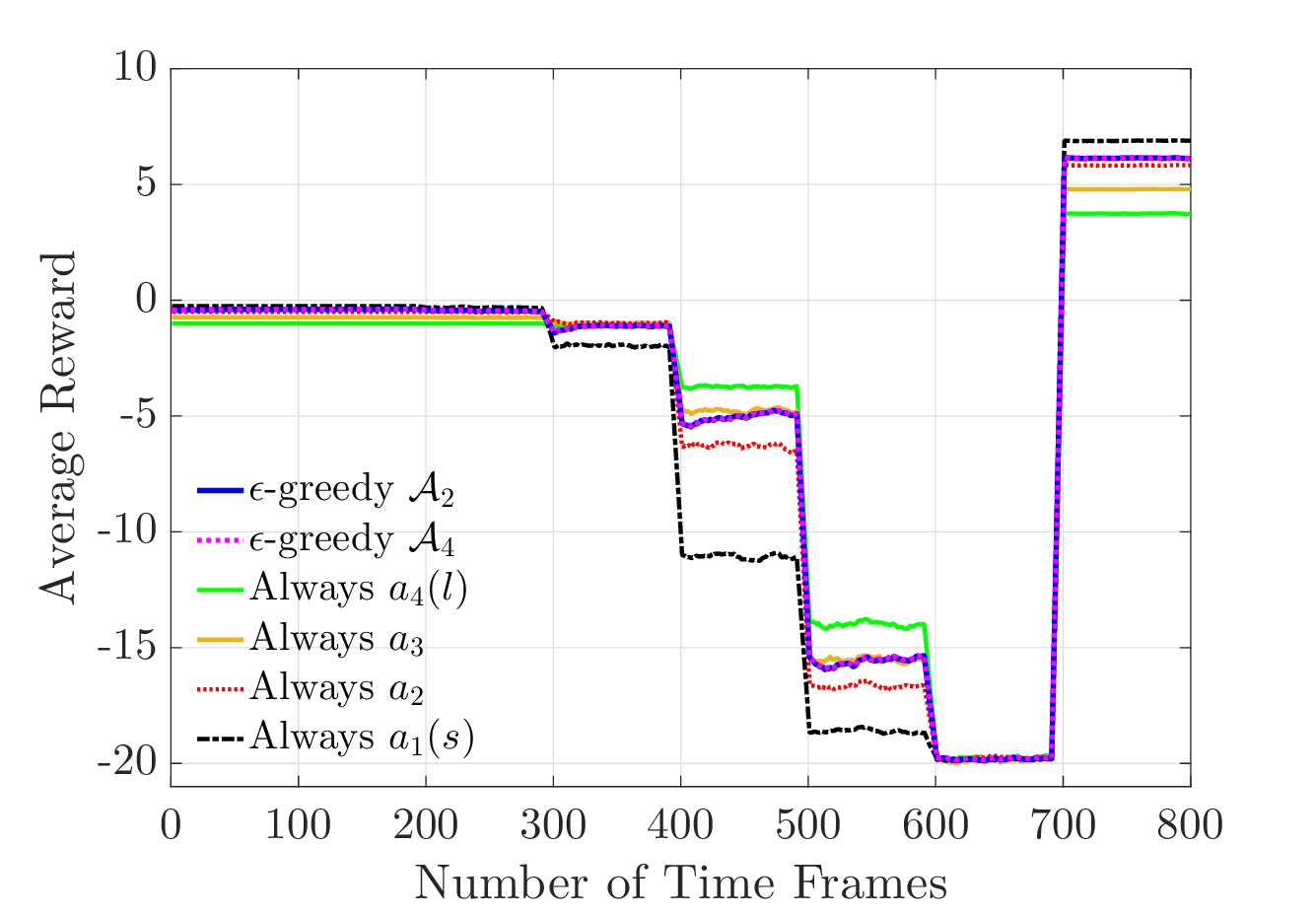}
\centering	
	\caption{{\fontsize{9pt}{9pt}\selectfont Smoothed average reward of policies with FLOM detector in different time frames.}
	{\label{fig12}} }
	\vspace{-0.7cm}
\end{figure}
Considering all frames, it is evident that the agent employing the $\epsilon$-greedy policy consistently outperforms, as its average reward remains consistently close to the optimal action across all sections.
   Table \ref{tab5} confirms the total superiority of $\epsilon$-greedy for selecting NAS-CNN with short and long sensing times. In this table, the mean of average reward on all frames and iterations for different policies has been reported. With a substantial gap, $\epsilon$-greedy selects the optimal action in a larger number of frames and receives far less punishment than the always short and GB policies.\par
  The proposed RL-based method for sensing time selection can be used for model-based detectors. In Fig. \ref{fig12}, the same simulation was used with FLOM detector where channel environment is more rapidly changing in comparison to the previous case. In this simulation, 800 time frames are divided to 8 sections with 100 time frames. In these sections, the channel conditions $(\mathsf{GSNR},H_i)$ are $(30 \mathrm{dB},H_1)$, $(25 \mathrm{dB},H_1)$, $(20 \mathrm{dB},H_1)$, $(15 \mathrm{dB},H_1)$, $(10 \mathrm{dB},H_1)$, $(5 \mathrm{dB},H_1)$, $(0 \mathrm{dB},H_1)$, and $(-,H_0)$, respectively. Here, we  employed $\epsilon$-greedy policy with $\epsilon=\alpha_{lr}=0.15$ and with two different action spaces. One action space is $\mathcal{A}_2$ with two actions used in Fig. \ref{fig11} as well, and the second one is $\mathcal{A}_4=\{a_1, a_2, a_3, a_4\}$ with four actions where $T^a_{sen}\in\{T^{a_1}_{sen}=8, T^{a_2}_{sen}=16, T^{a_3}_{sen}=24, T^{a_4}_{sen}=32\}$ \textmu s. This action space gives more options to the agent at the cost of estimation of four $Q(a)$ which requires more sampling. As shown in Fig. \ref{fig12} and Table \ref{tab5}, agents using $\epsilon$-greedy perform effectively and achieve larger mean average reward in the entire frames. For instance, $\epsilon$-greedy with $\mathcal{A}_4$ approximately yields a 20\% increase in the gained rewards compared to the Always $a_1$ policy. Comparing  both $\epsilon$-greedy policies, the agent with $\mathcal{A}_4$ very slightly outperforms the agent with action space $\mathcal{A}_2$. 
   
  
%
%
{\renewcommand{\arraystretch}{1.1}
\begin{table}[t!]
\fontsize{9pt}{10pt}\selectfont 
  \begin{center}
    \caption{{\fontsize{9pt}{9pt}\selectfont Performance of different policies in time frames of Fig. \ref{fig11} for NAS-CNN detectors and Fig. \ref{fig12}} for FLOM detector.\label{tab5}}

    \begin{tabular}{
  >{\centering}m{0.10\textwidth}|
 >{\centering}m{0.15\textwidth}|
    >{\centering\arraybackslash}m{0.15\textwidth}} 
  \toprule[1.5pt]

\textbf{Detector}&\textbf{Policy} &   \textbf{Mean of Average Reward} \\
\toprule[1.5pt]
\multirow{4}{*}{NAS-CNN}& Always $a_4 (l)$   & -5.1193\\
      \cline{2-3}
   &Always  $a_1 (s)$        & -4.8766 \\
      \cline{2-3}
    & GB    $\mathcal{A}_2$  & -4.8977\\
      \cline{2-3}
    &$\epsilon$-greedy   $\mathcal{A}_2$     & -4.3887\\
     \toprule[1.5pt]
    \multirow{6}{*}{FLOM}& Always  $a_4 (l)$   & -4.2335\\
      \cline{2-3}
      &Always     $a_3$      & -4.2598 \\
      \cline{2-3}
      &Always     $a_2$      & -4.2880 \\
      \cline{2-3}
   &Always    $a_1 (s)$      & -4.9361 \\
      \cline{2-3}
    & $\epsilon$-greedy   $\mathcal{A}_2$  & -3.9559\\
      \cline{2-3}
    &$\epsilon$-greedy   $\mathcal{A}_4$    & -3.9518\\
      \toprule[1.5pt]
    \end{tabular}
  \end{center}
  \vspace{-.7cm}
\end{table}
}

\vspace{-0.3cm}
\section{Conclusion}\label{conclusion}
In this work, we discussed the challenging problem of architecture selection for SS with CNN-based detectors. We developed a NAS method which uses Q-learning to systematically search different architectures for CNNs. The proposed NAS method was  utilized to suggest the best networks for three different datasets. We attempted to include important signal, noise, and channel models in these datasets in order to cover empirical conditions that  SU should face in the process of SS. Thanks to the effective selection of  architectures by the NAS method, the proposed NAS-CNNs showed their superior performance in comparison with several other state-of-the-art detectors in the literature. Moreover, an SU, equipped with various detectors with different sensing times, is able to dynamically select the sensing time and obtain a better performance. To achieve this, we regarded this selection problem as an MAB, and we used related RL policies. We also designed a rewarding approach to force the agent to consider the created interference, the achieved throughput and the consumed energy in the process of sensing time adjustment. Simulation results show that an SU with $\epsilon$-greedy policy for the selection of sensing time of detector based on the state of channel and the received signal strength, accurately avoids causing interference for PUs and also obtains higher throughput and energy conservation.


 \printbibliography

\end{document}